\documentclass[12pt,oneside,english]{amsart}
\usepackage[T1]{fontenc}
\usepackage[utf8]{inputenc}
\usepackage[letterpaper]{geometry}
\usepackage{booktabs}
\usepackage{amstext}
\usepackage{amsthm}
\usepackage{amssymb}
\usepackage{setspace}
\usepackage[authoryear]{natbib}
\usepackage{float}
\usepackage [displaymath,mathlines]{lineno}

\makeatletter

\numberwithin{equation}{section}
\numberwithin{figure}{section}
\theoremstyle{plain}

\usepackage{chngcntr}
\usepackage{apptools}
\AtAppendix{\counterwithin{lemma}{section}}

\theoremstyle{plain}

\def\X{\mathcal{X}}

\theoremstyle{plain}

\numberwithin{equation}{section}

\usepackage[within=none]{caption}

\usepackage{subfig}
\usepackage{babel}
\usepackage{amsmath}
\usepackage{mathtools}
\usepackage{bbm}
\usepackage{relsize}

\usepackage{algpseudocode}

\usepackage[ruled]{algorithm2e}

\usepackage{array}
\newcommand{\mockalph}[1]{}

\makeatother

\usepackage{babel}
\providecommand{\theoremname}{Theorem}

\begin{document}

\title[HMC with Energy Conserving Subsampling]{Hamiltonian Monte Carlo with Energy Conserving Subsampling}

\author{Khue-Dung Dang$^{1}$}
\thanks{Corresponding author: Matias Quiroz, Sydney NSW 2007, e-mail: quiroz.matias@gmail,com}
\thanks{$^1$UNSW Business School, School of Economics, University of New South Wales.}
\author{Matias Quiroz$^{2, 3}$}
\thanks{$^2$School of Mathematical and Physical Sciences, University of Technology Sydney.}
\thanks{$^3$Research Division, Sveriges Riksbank.}
\author{Robert Kohn$^{1}$}
\author{Minh-Ngoc Tran$^{4}$}
\thanks{$^4$Discipline of Business Analytics, University of Sydney.}
\author{Mattias Villani$^{5, 6}$}
\thanks{$^5$Division of Statistics and Machine Learning,  Link\"{o}ping University.}
\thanks{$^6$Department of Statistics, Stockholm University.}
\thanks{All authors are affiliated with ARC Centre of Excellence for Mathematical and Statistical Frontiers (ACEMS)}

\begin{abstract}
Hamiltonian Monte Carlo (HMC) samples efficiently from high-dimensional posterior distributions with proposed parameter draws obtained by iterating on a discretized version of the Hamiltonian dynamics. The iterations make HMC computationally costly, especially in problems with large datasets, since it is necessary to compute posterior densities and their derivatives with respect to the parameters. Naively computing the Hamiltonian dynamics on a subset of the data causes HMC to lose its key ability to generate distant parameter proposals with high acceptance probability. The key insight in our article is that efficient subsampling HMC for the parameters is possible if both the dynamics and the acceptance probability are computed from the same data subsample in each complete HMC iteration. We show that this is possible to do in a principled way in a HMC-within-Gibbs framework where the subsample is updated using a pseudo marginal MH step and the parameters are then updated using an HMC step, based on the current subsample. We show that our subsampling methods are fast and compare favorably to two popular sampling algorithms that utilize gradient estimates from data subsampling. We also explore the current limitations of subsampling HMC algorithms by varying the quality of the variance reducing control variates used in the estimators of the posterior density and its gradients.
\\
\noindent \textsc{Keywords}: Bayesian inference, Big Data, Markov chain Monte Carlo, Estimated likelihood, Stochastic gradient Hamiltonian Monte Carlo, Stochastic Gradient Langevin Dynamics.\newpage{}
\end{abstract}

\maketitle

\section{Introduction\label{sec:Introduction}}
Bayesian inference relies on computing expectations with respect to the posterior density of the model parameters given the data. The functional form of the posterior density often does not correspond to a known density and hence obtaining independent samples to compute the expectation by Monte Carlo integration is difficult, especially when the dimension of the model parameter is moderate to large. Markov Chain Monte Carlo (MCMC) is a generic sampling algorithm that produces correlated draws from the posterior density.

Metropolis-Hastings (MH) \citep{metropolis1953equation, hastings1970monte} is arguably the most popular MCMC algorithm. Its most common implementation uses a random walk proposal, in which a new sample is proposed based on the current state of the Markov chain. While Random walk MH is easy to implement, it explores the posterior very slowly in high-dimensional problems and gives highly correlated draws and imprecise estimators of posterior integrals. 

Hamiltonian Monte Carlo (HMC) \citep{duane1987hybrid} can produce distant proposals while maintaining a high acceptance probability \citep{neal2011mcmc, betancourt2017conceptual}. HMC augments the target posterior by adding fictitious momentum variables and carries out the sampling
on an extended target density. The extended target is proportional to the exponential
of a Hamiltonian function that describes the total energy of the system, which is the sum of the potential energy (negative log posterior) and the kinetic energy (negative log density of the momentum variables). The Hamiltonian dynamics describes how the total energy evolves through time.
One particularly interesting feature of the Hamiltonian is that it
conserves energy as time evolves, a property that is approximately
maintained even when the dynamics is approximated in discrete time.
Hence, a MH proposal obtained by simulating the dynamics has approximately the same value of the extended target density as
that of the current draw, resulting in a high acceptance probability,
even when the proposed draw is far from the current draw. This typically avoids the inherently slow exploration of the parameter space evident in random walk proposals \citep{betancourt2017conceptual}. HMC simulates the evolution of Hamiltonian dynamics for a given period of time. Up to a point, the longer the integration time the more effectively the dynamics explore the posterior distribution, while small integration times approach diffusive Langevin methods \citep{roberts1998optimal,roberts2002langevin}. In practice, the simulation of the dynamics is implemented with a numerical integrator. The more accurate the integrator, the larger the step size and the fewer total steps needed to simulate the dynamic for a chosen time. Smaller step sizes typically require larger integration times to achieve the same efficient exploration of the posterior distribution. In either case, HMC requires computing the gradient of the log-posterior in each step when simulating the dynamics, and in practice  a large number of steps may be performed. The extra computations are often worthwhile as HMC greatly improves the sampling efficiency of the generated samples compared to a proposal which does not use gradient information. Even with the improved performance of HMC, however, the cost can be too prohibitive for the limited computational resources available in a given application. Consequently, some algorithms use subsampling of the data to reduce the cost of computing the posterior density and its gradients. Unfortunately, naive subsampling methods are often problematic and in particular they compromise many of the features that give HMC its scalability \citep{betancourt2015fundamental}. In this article we introduce a subsampling HMC scheme that can tackle very large datasets and maintain the scalability.

Our article speeds up computation by using subsets of the data to compute both the dynamics and the subsequent MH correction performed when deciding to accept a proposal. More precisely, we propose a HMC-within-Gibbs algorithm that alternates i) sampling small subsets of data using a pseudo-marginal step and ii) sampling parameters using the HMC dynamics and the MH correction based on the current subset of the data. We will here focus on HMC algorithms where the Hamiltonian dynamics are used to generate a proposal which is subsequently accepted or rejected using a MH step, which we refer to as HMC. Extensions to other HMC algorithms that utilize the entire trajectory generated by the dynamics \citep{hoffman2014no,betancourt2017conceptual} are interesting future research direction discussed in Section \ref{sec:Conclusions}.

We propose two different subsampling versions of the HMC algorithm. In the first \textit{perturbed} approach we use a slightly biased likelihood estimator and show that the algorithm targets a perturbed posterior which gets close to the true posterior density as the subsample size and the number of observations of the dataset become large; see Section \ref{subsec:EfficientEstimatorsLikelihoodAndGradient} for details. The second approach uses an unbiased but possibly negative likelihood estimator which allows us to obtain simulation consistent estimators of the posterior expectation of any function of the parameters. However, this approach is harder to implement efficiently than the perturbed approach and is typically slower computationally. 

We compare our algorithms to Stochastic Gradient Langevin Dynamics \citep[SGLD]{welling2011bayesian} and Stochastic Gradient Hamiltonian Monte Carlo \citep[SG-HMC]{chen2014stochastic}, two of the most popular subsampling algorithms that utilize gradient information in machine learning. To make the comparison more challenging we implement both methods with control variates for improved performance \citep{baker2017control}. We demonstrate that our algorithms compare favorably to SGLD and SG-HMC. It is by now well known that all proposed subsampling MCMC and HMC algorithms need accurate control variates to lower the variance of the estimated posterior and gradients, and we explore the robustness of the algorithms when the control variates are degraded. 

The paper is organized as follows. Section \ref{sec:previous_research} reviews previous research and the methods we compare against. Section \ref{sec:Methodology} presents our methodology using a general likelihood estimator and argues that it circumvents the incompatibility of data subsampling and HMC raised in recent literature. Sections \ref{sec:Approx_HMCwG} and \ref{sec:Exact_HMCwG} present two algorithms based on two specific likelihood estimators. Finally, Section \ref{sec:Application} demonstrates the usefulness of the methods on two large datasets and compares with alternative approaches. This section also explores the limitation of subsampling approaches by experimenting with successively degrading the quality of the control variates used for variance reduction. Section \ref{sec:Conclusions} concludes and discusses future research.

\section{Related work}
\label{sec:previous_research}
\subsection*{High-dimensional MCMC for large datasets}
There has recently been a surge of interest in developing posterior simulation algorithms that scale with respect to both the number of observations $n$ and the number of parameters $d$. Since simulation methods have the ambitious goal of exploring all regions in parameter space with sizable probability mass, they naturally require many more iterations than posterior optimization algorithms. Posterior optimization is computationally attractive for big data, but does not quantify the posterior uncertainty, which is often a central task in science. Although there exist optimization-based methods that aim to approximate the entire posterior distribution, e.g. variational Bayes \citep{blei2017variational}, Laplace approximations \citep[Chapter 5]{bernardo2001bayesian} or integrated nested Laplace approximations \citep{rue2009approximate}, in practice it is nearly impossible to know how they perform without comparing the results to a posterior simulation method. It is thus important to develop posterior simulation methods that:\begin{itemize}
    \item[i)]remain computationally efficient when $n$ is large and 
    \item[ii)] explore the posterior distribution efficiently when $d$ is large.
\end{itemize}
Two distinct approaches exist to resolve i). The first is to utilize parallel computing by dividing the $n$ data observations into $K$ parts, performing independent posterior simulation on each of the $K$ subposteriors and subsequently merge the draws to represent the full data posterior. See, for example, \cite{scott2013bayes,neiswanger2013asymptotically,wang2013parallel,minsker2014scalable,nemeth2016merging}. The second approach, which is the focus of our article, is to work with subsamples of $m$ observations to estimate the full data posterior \citep{maclaurin2014firefly, korattikara2014austerity, bardenet2014towards, bardenet2015markov,  maire2018informed, bierkens2018zig, quiroz2018speeding, quiroz2016exact, quiroz2017delayed, gunawan2018subsampling} or its gradient \citep{welling2011bayesian, chen2014stochastic, ma2015complete, shang2015covariance, baker2017control}. The rest of this section reviews samplers which utilize gradient information about the posterior density.

The primary problem confronted in ii) is how to generate proposals which maintain a high acceptance probability and are also distant enough to avoid a highly persistent Markov chain for the model parameter. A useful approach is to simulate a discretized Langevin diffusion \citep{roberts1998optimal, nemeth2016particle} or, more generally, Hamilton's equations \citep{duane1987hybrid} and use the simulated draw as a proposal in the MH algorithm to correct for the bias introduced by the discretization \citep{neal2011mcmc}. HMC provides a solution to ii) \citep{neal2011mcmc, betancourt2017conceptual}, but when combined with i), the algorithm becomes computationally intractable since simulating the Hamiltonian dynamics requires a large number of costly gradient evaluations for every proposed parameter value. 

\subsection*{Subsampling HMC algorithms and related approaches}A computationally attractive way to accelerate HMC is to use a fixed subsample of the data to unbiasedly estimate the computationally costly gradients in each step of the discretized Hamiltonian trajectory, and skip the MH correction to avoid evaluating the posterior on the full data. \cite{betancourt2015fundamental} demonstrates that this simple strategy produces highly biased trajectories, where the bias depends upon the quality of the gradient estimator. Moreover, \cite{betancourt2015fundamental} shows that attempts to average out the bias by renewing the subsample in each step of the trajectory still perform poorly; see also the \emph{naive stochastic gradient HMC} in \cite{chen2014stochastic}. \cite{betancourt2015fundamental} illustrates that adding a MH correction step based on the full data to fully correct for the biased trajectories leads to a rapid deterioration of the acceptance probability of HMC as $d$ increases, and concludes that there is a fundamental incompatibility of HMC and data subsampling. As a remedy to the poor performance by the naive stochastic gradient HMC, \cite{chen2014stochastic} propose adding a friction term to the dynamics to correct for the bias. For the rest of our article, we refer to the method using the friction term in the dynamics as Stochastic Gradient Hamiltonian Monte Carlo (SG-HMC). \cite{chen2014stochastic} omit the MH correction step and the resulting bias in the posterior depends on the quality of the discretization of the continuous dynamics. Hence, in order to traverse the parameter space effectively, the integrator potentially needs a large number of steps to compensate for the small step size used when discretizing the dynamics, which may be very costly.

A different approach, but related in the sense that it uses an estimated gradient, is Stochastic Gradient Langevin Dynamics \citep[SGLD]{welling2011bayesian}. SGLD combines Stochastic Gradient Optimization \citep[SGO]{robbins1951stochastic} and Langevin dynamics \citep{roberts1998optimal}, by allowing the initial iterates to resemble a SGO and gradually traverse to Langevin Dynamics so as to not collapse to the posterior mode. SGLD avoids a costly MH correction by arguing that it is not needed because the discretization step size of the Langevin dynamics is decreased as a function of the iterates \citep{welling2011bayesian}. However, this decreases the rate of convergence of estimators based on its output to $R^{-1/3}$, where $R$ is the number of samples from the posterior \citep{teh2016consistency}, as opposed to $R^{-1/2}$ for MCMC \citep{roberts2004general}, and in particular HMC. Practical implementations of SGLD use a sequence of step sizes that does not decrease to zero in the limit, and \citet{vollmer2015non} show that the posterior approximated by SGLD can then be quite far from the true posterior; see also \cite{Brosse2018}. \cite{bardenet2015markov} also demonstrate that SGLD can be accurate for the posterior mode, but gives a poor approximation of the full posterior distribution on a toy example with $d=2$ and highly redundant data, i.e. superfluous amounts of data in relation to the complexity of the model. Recently, \cite{dubey2016variance} improve SGLD using control variates, see also \cite{baker2017control} who, in addition, use control variates in the SG-HMC algorithm proposed in \cite{chen2014stochastic}. We implement both SGLD and SG-HMC with highly efficient control variates when comparing them to our method. We also implement the methods without control variates as it has been shown that sometimes variance reduction may be detrimental \citep{pmlr-v80-chatterji18a}.

All the problems discussed above with subsampling in HMC and related algorithms stem from the fact that subsampling disconnects the Hamiltonian from its own dynamics. This disconnect causes HMC proposals to lose their energy conserving property and their attractive ability to sample efficiently in high dimensions. The next section presents a new energy conserving approach to subsampling in HMC that keeps the connection intact by estimating both the Hamiltonian and its corresponding dynamics from the same subsample. By updating the subsample in a separate pseudo-marginal step we make sure that the HMC algorithm still targets the posterior based on all data. Put differently, our new approach creates a Hamiltonian system with corresponding dynamics for a given subset of data. This allows for the scalability of HMC to be maintained for each subsample, as is demonstrated in Section \ref{subsec:scalability}.
  
\section{Energy conserving subsampling in HMC}
\label{sec:Methodology}
This section presents our new approach to subsampling in HMC and discusses why our approach avoids the pitfalls described in \cite{betancourt2015fundamental}. In order to not distract from the main ideas, we first present our approach for a general unbiased and almost surely positive likelihood estimator from data subsampling. Sections \ref{sec:Approx_HMCwG} and \ref{sec:Exact_HMCwG} then present practical algorithms based on likelihood estimators proposed in \citet{quiroz2018speeding} and \citet{quiroz2016exact}, respectively.

\subsection{Pseudo-marginal MCMC}
Denote the model parameter vector by $\theta \in \Theta \subset \mathbb{R}^{d}$, where $\mathbb{R}^{d}$ is the space of $d$-dimensional real vectors, and let $\pi(\theta)$ be its posterior density given a dataset $y$ with $n$ observations. We first briefly describe pseudo-marginal MCMC \citep{andrieu2009pseudo}, which serves as inspiration for one building block in our subsampling HMC approach. Pseudo-marginal algorithms  target the augmented posterior
\begin{equation}
\overline{\pi}_{m}(\theta,u) \propto \widehat{L}_m(\theta) p_{\Theta}(\theta)p_{U}(u),
\label{eq:AugmentedPosterior}
\end{equation}
where $\widehat{L}_{m}(\theta)$ is an unbiased and non-negative estimator of the likelihood $L(\theta)$ based on $m$ auxiliary variables, $u$, with density $p_U(u)$. In the particular application to subsampling, $u\in \{1,\dots,n\}^m$, contains the indices for the data observations used in estimating the likelihood and $m$ denotes the subsample size, see Section \ref{subsec:logl_and_grad_logl_estimators} for details. Note that $\widehat{L}_m$ and any quantity included in its definition depend on $n$, but this is suppressed in the notation for conciseness. We can now design an MCMC chain to sample $\theta$ and $u$ jointly from \eqref{eq:AugmentedPosterior} and, since $\widehat{L}_m(\theta)$ is unbiased, the $\theta$ iterates are samples from $\pi(\theta)$.

The choice of $m$ is crucial for pseudo-marginal methods. An $m$ that is too small results in a noisy likelihood estimator and the Markov chain may get stuck due to severely overestimating the likelihood at the current draw, subsequently rejecting nearly all proposals. Conversely, taking $m$ too large wastes useful computational resources. A natural aim is to choose an $m$ that minimizes the Computational Time (CT) needed to generate the equivalent of a single independent draw from the posterior,  with
\begin{equation}\label{eq:CT}
\mathrm{CT} \coloneqq \mathrm{IF}\times  \text{Total number of density and gradient evaluations},
\end{equation}
where the Inefficiency Factor (IF) is proportional to the asymptotic variance when estimating a posterior functional based on the MCMC output, and is interpreted as the number of samples needed to obtain the equivalent of a single independent sample. The second term is proportional to the computing time for a single draw, a measure that is independent of the implementation. Starting with \cite{pitt2012some}, there is a large literature showing that CT is minimized by an $m$ that targets a variance of the log of the likelihood estimator around 1 \citep{doucet2015efficient,  sherlock2015pseudo, tran2016block, deligiannidis2015correlated, schmon2018large}. Recent developments in pseudo-marginal methods induce dependence in the auxiliary variables $u$ over the MCMC iterations such that the likelihood estimates over the iterations become dependent \citep{deligiannidis2015correlated}. This makes it possible to tolerate a substantially larger variance of the likelihood estimator, and hence smaller subsamples in our context. We follow \cite{tran2016block} and induce the dependence over the iterations by partioning the $u$'s into blocks and only update a subset of the blocks in each iteration. The optimal subsample size $m$ is then obtained by targeting a certain value for the conditional variance of the likelihood estimator for a given induced correlation $\rho$ \citep{tran2016block}.

\subsection{An energy conserving HMC-within-Gibbs framework}
\label{sec:main_idea}
Following the standard HMC algorithm, our subsampling HMC algorithm introduces a fictitious continuous momentum vector $\vec{p} \in \mathcal{P} \subset \mathbb{R}^d$ of the same dimension as the continuous parameter vector $\theta$. The extended target in \eqref{eq:AugmentedPosterior} is then further augmented by $\vec{p}$ to
\begin{equation}
\overline{\pi}_{m}(\theta,\vec{p},u) \propto \exp\left(-\widehat{\mathcal{H}}(\theta, \vec{p}) \right) p_U(u), \quad \widehat{\mathcal{H}}(\theta, \vec{p}) = \widehat{\mathcal{U}}(\theta) + \mathcal{K}(\vec{p})
\label{eq:AugmentedPosterior2}
\end{equation}
with
\begin{equation}
    \widehat{\mathcal{U}}(\theta) =  -\log \widehat{L}_{m}(\theta) - \log p_{\Theta}(\theta)\quad \text{and } \mathcal{K}(\vec{p}) =\frac{1}{2}\vec{p}^{\, \, \prime}M^{-1}\vec{p},
    \label{eq:energies}
\end{equation}
where $M$ is a symmetric positive-definite matrix. In \eqref{eq:AugmentedPosterior2}
we assume that the Hamiltonian $\widehat{\mathcal{H}}$ is separable. We propose a HMC-within-Gibbs method to sample from \eqref{eq:AugmentedPosterior2}, alternating sampling from
\begin{enumerate}
\item $u | \theta, \vec{p}, y$ - Pseudo-marginal MH update (Section \ref{GibbsStep1})
\item $\theta, \vec{p} \, | u, y$ - HMC update given $u$ from Step 1 (Section \ref{GibbsStep2}).
\end{enumerate}
 This scheme has \eqref{eq:AugmentedPosterior2} as its invariant distribution. Integrating out the momentum variables yields $\overline{\pi}_{m}(\theta,u)$ in \eqref{eq:AugmentedPosterior} and, further integrating out $u$, yields $\pi(\theta)$ if the likelihood estimator $\widehat{L}_m(\theta)$ is unbiased. \cite{lindsten2016pseudo} propose the related pseudo-marginal HMC sampler, in which a momentum vector is also introduced for the auxiliary variables $u$. That scheme, however, is not applicable here as the pseudo-marginal variables we employ are discrete and not amenable to Hamiltonian dynamics themselves.

The next subsections describe in detail the two updates of our algorithm and explain why our approach does not compromise the Hamiltonian flow.

\subsection{Updating the data subset}
\label{GibbsStep1}
Given $\theta^{(j-1)}, \vec{p}^{\,\,(j-1)}$ and $u^{(j-1)}$, at iteration $j$  we propose $u^{\prime} \sim p_U(u)$ and set $u^{(j)} = u^{\prime}$ with probability
\begin{equation}
\alpha_{u} = \min\left\{1, \frac{\widehat{L}_m(\theta^{(j-1)};u^\prime)}{\widehat{L}_m(\theta^{(j-1)};u^{(j-1)})} \right\},
 \label{eq:accProb_u}
\end{equation}
where the notation emphasizes that the estimators are based on different data subsets. If the proposal is rejected we set $u^{(j)} = u^{(j-1)}$.

Since $u^{\prime}$ is proposed independently of $u^{(j-1)}$, the log of the ratio in \eqref{eq:accProb_u} can be highly variable, possibly getting the sampler stuck when the numerator is significantly overestimated. To prevent this, we implement the block update of $u$ for data subsampling in \cite{quiroz2018speeding, quiroz2016exact} with $G$ blocks, which gives a correlation $\rho$ of roughly $1 - 1/G$ between the $\log \widehat{L}_{m}$ at the current and proposed draws \citep{tran2016block, quiroz2016exact}. Setting $G=100$, gives $\rho \approx 0.99$, which helps the chain to mix well.

\subsection{Updating the parameters}
\label{GibbsStep2}
Given $u^{(j)}$, we use Hamilton's equations
\begin{equation}
\frac{d\theta_l}{dt}  =  \frac{\partial \widehat{\mathcal{H}}(\theta,\vec{p})}{\partial \vec{p}_l}, \quad \frac{d\vec{p}_l}{dt}  =  -\frac{\partial \widehat{\mathcal{H}}(\theta,\vec{p})}{\partial \theta_l}, \quad l =1, \dots , d, 
\label{eq:modifiedDynamics}
\end{equation}
%
to propose $\theta$ and $\vec{p}$. Note that this trajectory follows the Hamiltonian flow for $\widehat{\mathcal{H}}$ viewed as a function of $\theta$ and $\vec{p}$ for a given data subset selected by $u^{(j)}$, since $u^{(j)}$ is fixed through time $t$. We obtain the proposal as in standard $\mathrm{HMC}$, using a leapfrog integrator with integration time $\epsilon L$, but with $\widehat{\mathcal{H}}$ in place of $\mathcal{H}$. Specifically, at iteration $j$, given the data subset $u^{(j)}$, if the leapfrog integrator starts at $(\theta^{(j-1)}, \vec{p}_0)$ with $\vec{p}_0 \sim \mathcal{K}(\vec{p})$ and ends at $(\theta_{L}, -\vec{p}_L)$, we let $(\theta^{(j)}, \vec{p}^{\,\,(j)}) = (\theta_L, -\vec{p}_L)$ with probability 
\begin{equation}
 \label{eq:accProb_theta_and_p}
\alpha_{\theta, \vec{p}} =   \min\left\{1,\exp\left(-\widehat{\mathcal{H}}(\theta_L, -\vec{p}_L) + \widehat{\mathcal{H}}(\theta^{(j-1)}, \vec{p}_0)\right)\right\},
\end{equation} 
with $\widehat{\mathcal{H}}$ in \eqref{eq:AugmentedPosterior2}. If $(\theta_{L}, -\vec{p}_L)$ is rejected,  we set $(\theta^{(j)}, \vec{p}^{\,\,(j)}) =  (\theta^{(j-1)}, \vec{p}_0)$. In practice, it is unnecessary to store the sampled momentum.

Using the terminology  in \cite{betancourt2015fundamental}, we can think of the dynamics in \eqref{eq:modifiedDynamics} as generating a trajectory following a \textit{modified level set} ($\widehat{\mathcal{H}}$), as opposed to the \textit{exact level set} obtained using dynamics that do not subsample the data ($\mathcal{H}$). A key property of our framework is that the \emph{same} estimate $\widehat{\mathcal{H}}$ is used in generating the discretized leapfrog trajectory as in the acceptance probability in \eqref{eq:accProb_theta_and_p}. The connection between the modified level set $\widehat{\mathcal{H}}$ and its dynamics is kept intact, and thus the original energy conserving property of HMC remains, even for distant proposals. We therefore name our algorithm Hamiltonian Monte Carlo with Energy Conserving Subsampling (HMC-ECS). Note that energy conservation is only possible because of the pseudo-marginal mechanism where we update the subsample in each Gibbs iteration, thereby guaranteeing that our samples are from the target posterior based on all the data.

\cite{betancourt2015fundamental} illustrates the problems with using $\widehat{\mathcal{H}}$ for the dynamics, but $\mathcal{H}$ in the acceptance probability. Given a sensible step length $\epsilon$, discretizing the Hamiltonian with a symplectic integrator introduces an error of $O(\epsilon^2)$ \citep{neal2011mcmc} relative to the modified level set and hence the discretization error is very small. \cite{betancourt2015fundamental} notes that the modified level set and the discretized trajectory based on it might be very far from the exact level set, resulting in low acceptance probabilities no matter how small $\epsilon$ is. SG-HMC \citep{chen2014stochastic} deliberately circumvents the disconnect problem by generating proposals from the trajectories based on a modified Hamiltonian, but skip the rejection step. The disadvantage of SG-HMC is therefore that the bias in the targeted posterior now grows with the step length $\epsilon$. Keeping $\epsilon$ small makes SG-HMC very computationally demanding since a very large number of leapfrog steps are needed for distant proposals. In contrast, the dynamics of HMC-ECS target the subsampled Hamiltonian, and so maintain a high acceptance probability even for a large $\epsilon$. The bias introduced by the subsampling is then confined to the pseudo-marginal step, which is chosen so that the bias is very small as our theoretical analysis below shows. 

Algorithm \ref{Alg:HMC-wG} shows one iteration of our proposed $\mathrm{HMC\mbox{-}ECS}$ algorithm based on the leapfrog integrator using the estimated likelihood $\widehat{L}_m(\theta)$. The next two sections consider previously proposed likelihood estimators and show how we use them in $\mathrm{HMC\text{-}ECS}$. 

\begin{algorithm}
	\caption{One iteration of $\mathrm{HMC\mbox{-}ECS}$.} 
	\SetKwInOut{Input}{Input}
	\vspace{1mm}
	\Input {Current position $u^{(j-1)}$, $\theta^{(j-1)}$, stepsize $\epsilon$ and integrating time $\epsilon L$}
	\vspace{1mm}
	Propose $u' \sim p_U(u)$
		\vspace{1mm}

	Set $u^{(j)} \leftarrow u'$ with probability 
	$$\alpha_{u} = \min\left\{1, \frac{\widehat{L}_m(\theta^{(j-1)};u^\prime)}{\widehat{L}_m(\theta^{(j-1)};u^{(j-1)})} \right\},
$$ else $u^{(j)} \leftarrow u^{(j-1)}$\vspace{1mm} \\ \textbf{Given} $u^{(j)}$ : \\
			\vspace{1mm}
	$\vec{p}_0 \sim \mathcal{K}(\vec{p});\, \theta_0 \leftarrow \theta^{(j-1)}; \, \widehat{\mathcal{H}}_0 \leftarrow \widehat{\mathcal{H}}(\theta^{(j-1)}, \vec{p}_0)$
				\vspace{1mm}

	$\vec{p}_0 \leftarrow \vec{p}_0 - \frac{\epsilon}{2} \nabla_{\theta} \widehat{\mathcal{U}}(\theta_0)$
					\vspace{1mm}

	\For{$l = 1$ \KwTo $L$} {
						\vspace{1mm}

		$\theta_l \leftarrow \theta_{l-1} + \epsilon M^{-1}\vec{p}_{l-1}$ \\
							\vspace{1mm}

		\lIf{$i<L$} {$\vec{p}_l \leftarrow \vec{p}_{l-1} - \epsilon \nabla_{\theta} \widehat{\mathcal{U}}(\theta_l)$}	
						\vspace{1mm}

		\lElse {$\vec{p}_L \leftarrow \vec{p}_{L-1} - \frac{\epsilon}{2} \nabla_{\theta} \widehat{\mathcal{U}}(\theta_l)$} 					\vspace{1mm}

	}
	$\vec{p}_L \leftarrow -\vec{p}_L$ \\
		\vspace{1mm}
	$\widehat{\mathcal{H}}_L \leftarrow \widehat{\mathcal{H}}(\theta_L, \vec{p}_L)$
	
	\vspace{1mm}
	Set $(\theta^{(j)}, \vec{p}^{\,\,^{(j)}}) \leftarrow (\theta_L ,\vec{p}_L)$ with probability $$\alpha_{\theta, \vec{p}} =   \min\left\{1,\exp\left(-\widehat{\mathcal{H}}(\theta_L, -\vec{p}_L) + \widehat{\mathcal{H}}(\theta^{(j-1)}, \vec{p}_0)\right)\right\},$$
	else $(\theta^{(j)}, \vec{p}^{\,\,^{(j)}}) \leftarrow (\theta^{(j-1)}, \vec{p}_0) $ \\	\vspace{1mm}

	\textbf{Output:} $u^{(j)}$, $\theta^{(j)}$ 

\label{Alg:HMC-wG}

\end{algorithm}

\newpage
\section{Perturbed HMC-ECS}
\label{sec:Approx_HMCwG}
\subsection{Efficient estimators of the log-likelihood}
\label{subsec:logl_and_grad_logl_estimators}
\cite{quiroz2018speeding} propose sampling $m$ observations with replacement and estimate an additive log-likelihood $$\ell(\theta)=\log L(\theta)=\sum_{k=1}^n \ell_k(\theta), \quad \ell_k(\theta) = \log p(y_k|\theta) $$ by the unbiased difference estimator 
\begin{equation}
\widehat{\ell}_{m}(\theta)=\sum_{k=1}^{n} q_k(\theta) + \widehat{d}_m(\theta), \label{eq:difference_estimator_logl}
\end{equation}
where
\begin{equation}
    \widehat{d}_m(\theta) =  \frac{1}{m}\sum_{i=1}^m \frac{\ell_{u_i}(\theta) - q_{u_i}(\theta)}{\omega_{u_i}(\theta)}, \quad u_i \in \{1, \dots, n\} \text{ iid with} \Pr(u_i = k)=\omega_k,
    \label{eq:dhat}
\end{equation}
and $q_{k}(\theta)$ are control variates. We continue to suppress dependence on $n$ in the notation for many quantities introduced in this section. If the $q_k(\theta)$ approximate the $\ell_k(\theta)$ reasonably well, then we obtain an efficient estimator by taking $\omega_k=1/n$ for all $k$. \cite{quiroz2018speeding} estimate $\sigma^2(\theta) = \mathrm{V}\left[\widehat{\ell}_{m}(\theta)\right]$ by 
\begin{equation}
\widehat{\sigma}_{m}^2(\theta) = \frac{n^2}{m^2} \sum_{i=1}^m \left(d_{u_i}(\theta) - \overline{d}_u(\theta)\right)^2, \quad \text{with } d_{u_i}(\theta) =\ell_{u_i}(\theta) - q_{u_i}(\theta)
\label{eq:variance_estimate}
\end{equation}
where $\overline{d}_u$ denotes the mean of the $d_{u_i}$ in the sample $u = (u_1, \dots, u_m)$.

To obtain efficient control variates, \cite{quiroz2018speeding} follow \cite{bardenet2015markov} and let $q_k(\theta)$ be a second order Taylor approximation around a fixed central value $\theta^{\star}$,
\begin{equation}
q_k(\theta) = \ell_k(\theta^{\star}) + \nabla_{\theta} \ell_k(\theta^{\star})^{\top}(\theta - \theta^{\star}) + \frac{1}{2} (\theta - \theta^{\star})^{\top} H_k(\theta^{\star})(\theta - \theta^{\star}), \quad H_k(\theta^{\star}) \coloneqq \nabla_{\theta}  \nabla_{\theta}^{\top} \ell_k(\theta^{\star}). \label{eq:TaylorProxy_logl}
\end{equation}
After processing the full data once before the MCMC to compute simple summary statistics, $\sum_{k=1}^{n} q_k(\theta)$ can be computed in $O(1)$ time \citep{bardenet2015markov}. 
\subsection{Efficient estimators of the likelihood and its gradient}
\label{subsec:EfficientEstimatorsLikelihoodAndGradient}
\cite{quiroz2018speeding} use the likelihood estimator
\begin{equation}
    \widehat{L}_m(\theta) = \exp\left(\widehat{\ell}_m(\theta)-\frac{1}{2}\widehat{\sigma}^2_m(\theta)\right)
    \label{eq:approximate_estimator}
\end{equation}
first proposed by \cite{ceperley1999penalty} and \cite{nicholls2012coupled}. The motivation for this estimator is that it is unbiased for the likelihood if i) $\widehat{\ell}_m(\theta) \sim \mathcal{N}(\ell(\theta),\sigma^2(\theta))$ (justified by the central limit theorem) and ii) $\widehat{\sigma}^2_m(\theta)$ in \eqref{eq:approximate_estimator} is replaced by the population quantity $\sigma^2(\theta)$. However, $\sigma^2(\theta)$ is not available in practice and the estimator in  \eqref{eq:approximate_estimator} is biased. This bias makes the MCMC algorithm in \cite{quiroz2018speeding} target a slightly perturbed posterior $\overline{\pi}_{m}(\theta)$. Assuming that the expansion point $\theta^\star$ in the control variates is the posterior mode based on all the data, \cite{quiroz2018speeding} prove that
\begin{equation}
    \int_\Theta{\left|\overline{\pi}_{m}(\theta)-\pi(\theta)\right|}d\theta =O\left(\frac{1}{nm^{2}}\right) \text{ and } \left|{\mathrm{E}_{\overline{\pi}_{m}}[h(\theta)]-\mathrm{E}_{\pi}[h(\theta)]}\right|= O\left(\frac{1}{nm^{2}}\right),\label{JASAratesIdeal} 
\end{equation}where $\overline{\pi}_m(\theta) = \mathrm{E}_{p_U}\left[\widehat{L}_m(\theta)\right]$ is the perturbed marginal for $\theta$ when using the likelihood estimator in \eqref{eq:approximate_estimator}. These results carry over to our Hamiltonian approach straightforwardly as we obtain the augmented target in \cite{quiroz2018speeding} after integrating out the momentum in \eqref{eq:AugmentedPosterior2} and using \eqref{eq:approximate_estimator}. Hence, the $\theta$ iterates from HMC-ECS converge to a perturbed posterior which may get arbitrarily close to the true posterior by increasing the subsample size $m$, or by increasing $n$ and letting $m = O(n^{\nu})$ for some $\nu > 0$. For example, if $\nu = 1/2$, then the above orders are $O(1/n^2)$ with respect to $n$. However, this extremely rapidly vanishing perturbation is usually not practically attainable since the result in \eqref{JASAratesIdeal} assumes that $\theta^\star$ is the posterior mode based on \emph{all} data. Corollary 1 in \cite{quiroz2018speeding} proves rates under the more realistic assumption that $\theta^\star$ is the posterior mode based on a fixed subset of $\tilde n \ll n$ observations. If, for example, $\tilde n = O(\sqrt{n})$ then the rates in \eqref{JASAratesIdeal} become $O(1/\sqrt{n})$. Importantly, the optimal subsample size in this case becomes $m=O(\sqrt{n})$, which shows that HMC-ECS scales well with the size of the data. See \cite{quiroz2018speeding} for suggestions on how to get closer to the rates in \eqref{JASAratesIdeal} in a computationally tractable way.

In addition to estimating the likelihood and the log-likelihood, our Hamiltonian approach also needs to estimate a gradient. It is straightforward to modify \eqref{eq:difference_estimator_logl} to instead provide an unbiased estimator of the gradient of the log-likelihood. With $\omega_k = 1/n$ and $\nabla_{\theta} q_k(\theta)= \nabla_{\theta} \ell_k(\theta^{\star}) + H_k(\theta^{\star}) (\theta - \theta^{\star})$,
\begin{equation}
\nabla_{\theta} \widehat{\ell}_{m}(\theta)= A(\theta^{\star}) + B(\theta^{\star})(\theta - \theta^{\star}) + \frac{n}{m} \sum_{i=1}^m \left(\nabla_{\theta}\ell_{u_i}(\theta) - \nabla_{\theta}q_{u_i}(\theta)\right),  \label{eq:difference_estimator_grad_logl}
\end{equation}
where $$\quad A(\theta^{\star}) \coloneqq \sum_{k=1}^n \nabla_{\theta} \ell_k(\theta^{\star}) \in \mathbb{R}^{d} \quad \text{and } B(\theta^{\star}) \coloneqq \sum_{k=1}^n H_k(\theta^{\star}) \in \mathbb{R}^{d \times d}$$
are obtained at the cost of computing over the full dataset once before the MCMC since $\theta^{\star}$ is fixed. It is also straightforward to compute $\nabla_{\theta} \widehat{\sigma}_m^2(\theta)$ using \eqref{eq:variance_estimate} with this choice of control variate.

It is important to note that the perturbation in the targeted posterior in perturbed HMC-ECS is independent of the step length in the leapfrog iterations, $\epsilon$. Hence, we can generate distant proposals from a small number of leapfrog steps without increasing the posterior bias.

\section{Signed HMC-ECS}
We now present an alternative HMC-ECS algorithm based on the Block-Poisson estimator in \cite{quiroz2016exact}. This algorithm gives simulation consistent estimates of expectations with respect to the true posterior density without any perturbation. 
\label{sec:Exact_HMCwG}
\subsection{The block-Poisson estimator}
\cite{quiroz2016exact} propose the block-Poisson estimator,
formed by sampling $\X_l \sim \mathrm{Pois}(1)$ for $l = 1, \dots \lambda,$ and computing $\widehat{d}_m^{\,\,(h,l)}$, $h =1, \dots, \X_l,$ using \eqref{eq:dhat} based on a mini-batch sample size $m$, and then estimate the likelihood by
\begin{eqnarray} \label{eq:PoissonEstimator}
	\widehat{L}_m(\theta) = \exp\left(\sum_{k=1}^n  q_k(\theta) \right) \prod_{l=1}^{\lambda}\xi_l, \, \hspace{0.5cm}\xi_l = \exp\left(\frac{a+\lambda}{\lambda}\right) \prod_{h=1}^{\X_l}\left(\frac{\widehat{d}_m^{\,\,(h,l)}(\theta)-a}{\lambda}\right), 
\end{eqnarray}
where $\lambda$ is a positive integer, $a$ is a real number and  $\xi_l = \exp\left((a + \lambda)/\lambda \right)$  if $\X_l=0$. $\mathrm{Pois}(1)$ denotes the Poisson distribution with mean 1. Note that the total subsample size $m \lambda \X_l$ is random with mean $m\lambda$ in a given MCMC iteration.

Since $\widehat{L}_m(\theta)$ is unbiased for $L(\theta)$, defining the augmented density as in \eqref{eq:AugmentedPosterior2} gives $$\int_U \int_{\mathcal{P}}  \overline{\pi}_{m}(\theta, \vec{p}, u)d\vec{p}du = \pi(\theta).$$ Hence, if \eqref{eq:AugmentedPosterior2} is a proper density using \eqref{eq:PoissonEstimator}, we obtain samples from the desired marginal for $\theta$. However, \eqref{eq:AugmentedPosterior2} is a proper density only if $\tau := \Pr\left(\widehat{L}_m(\theta) \geq 0\right)=1$,  which requires that $a$ in \eqref{eq:PoissonEstimator} is a lower bound of $\widehat{d}_{m}^{\,\,(h,l)}$ \citep{jacob2015nonnegative} which results in a prohibitively costly estimator \citep{quiroz2016exact}. Instead, we follow \cite{quiroz2016exact} who use the approach of \cite{lyne2015russian} for exact inference on an expectation of an arbitrary function $\psi(\theta)$ with respect to $\pi(\theta)$. For our Hamiltonian approach, this entails defining a proper augmented density, 
\begin{equation}\label{eq:AugmentedPosteriorExactApproach}
\widetilde{\pi}_m(\theta, \vec{p}, u) \propto \left|\widehat{L}_m(\theta)  \right| p_{\Theta}(\theta)p_{U}(u)\exp\left(-\mathcal{K}(\vec{p})\right),
\end{equation}
and writing
\begin{equation}
\mathrm{E}_{\pi}[\psi]   =  \frac{\int_\Theta  \psi(\theta)L(\theta)p_\Theta(\theta)d\theta}{\int_\Theta L(\theta) p_\Theta(\theta)d\theta} 
  =  \frac{ \int_{\mathcal{U}} \int_{\mathcal{P}} \int_\Theta  \psi(\theta)S(\theta,u)\widetilde{\pi}(\theta, \vec{p}, u)d\theta d\vec{p} du}{ \int_{\mathcal{U}} \int_{\mathcal{P}} \int_\Theta S(\theta,u)\widetilde{\pi}(\theta, \vec{p}, u)d\theta d\vec{p} du} 
  = \frac{\mathrm{E}_{\widetilde \pi}[\psi S]}{\mathrm{E}_{\widetilde \pi}[S]},\label{eq:EstimationOfFunctional}
\end{equation}
where $S(\theta, u) = \mathrm{sign}\left(\widehat{L}_m(\theta)\right)$ and $\mathrm{sign}(\cdot) = 1$ if  $\widehat{L}_m(\theta) \geq 0$ and $\mathrm{sign}(\cdot)=-1$ otherwise. 
Equation \eqref{eq:EstimationOfFunctional} suggests running the HMC-ECS sampler outlined in Section \ref{sec:main_idea} on the target \eqref{eq:AugmentedPosteriorExactApproach}, and then estimating \eqref{eq:EstimationOfFunctional} by 
\begin{equation*}
\widehat{I}_R = \frac{\sum_{j = 1}^R \psi(\theta^{(j)})s^{(j)}}{\sum_{j = 1}^R  s^{(j)}},
\end{equation*}
where $s^{(j)}$ is the sign of the estimate at the $j$th iteration. We follow \cite{quiroz2016exact} and use the term \emph{signed PM} for any pseudo-marginal algorithm that uses the technique in \cite{lyne2015russian} where a pseudo-marginal sampler is run on the absolute value of the estimated posterior and subsequently sign-corrected by importance sampling. Similarly, we call the algorithm described in this section \emph{signed HMC-ECS}.

The block-Poisson estimator in \eqref{eq:PoissonEstimator} has more tuning parameters than the estimator in \eqref{eq:approximate_estimator}. \cite{quiroz2016exact} extend the optimal tuning approach in \cite{pitt2012some} to the signed pseudo-marginal algorithm with the block-Poisson estimator. The Computational Time (CT) measure in \eqref{eq:CT} now becomes
\begin{equation}\label{eq:CTSigned}
\mathrm{CT} \coloneqq \frac{\mathrm{IF}}{(2\tau-1)^2} \times  \text{Total number of density and gradient evaluations},
\end{equation}
where $\tau := \Pr\left(\widehat{L}_m(\theta) \geq 0\right)$. \cite{quiroz2016exact} derive analytical expressions for both $\mathrm{V}[\log|\widehat{L}_m|]$ and $\mathrm{Pr}(\widehat{L}_m\geq0)$ needed to optimize CT. The fact that \cite{quiroz2016exact} take $\mathrm{Pr}(\widehat{L}_m\geq0)$ into account when tuning the algorithm avoids the instability from changing signs in signed PMMH. \cite{quiroz2016exact} also consider optimal tuning when correlating the estimators of the log of the likelihood at the current and proposed values of the MCMC. This correlation, $\rho$, is achieved by only updating $u$ for a subset of the products in \eqref{eq:PoissonEstimator} in each iteration, keeping the others fixed. \cite{quiroz2016exact} show that if $u$ is updated in $\kappa$ products, then  $\rho \approx 1 - \kappa /\lambda$.

\section{Applications\label{sec:Application}}
\subsection{Model}
We consider the logistic regression
\begin{eqnarray*}
p(y_{k}|x_{k},\theta) & = & \left(\frac{1}{1+\exp(-x_{k}^{\top}\theta)}\right)^{y_{k}}\left(\frac{1}{1+\exp(x_{k}^{\top}\theta)}\right)^{1-y_{k}},\quad \text{with } p_{\Theta}(\theta)=\mathcal{N}(\theta|0,\lambda_\theta^{-2}I ),
\end{eqnarray*}
where $\lambda_\theta$ is a global shrinkage factor which we treat as constant for simplicity. We estimate the model on two large datasets described below.

\subsection{Competing algorithms and performance measure}
We compare the performance of both the perturbed and the signed $\mathrm{HMC\mbox{-}ECS}$ algorithms against SGLD and SG-HMC. All subsampling methods use the same control variates based on a second order Taylor expansion for comparability; see Section \ref{subsec:Limitations} for experiments with control variates based on lower order Taylor expansions. The expansion point $\theta^\star$ is unique to each experiment and is discussed later.

Following \cite{vollmer2015non,baker2017control}, we implement SGLD using a fixed small step size $\epsilon$ instead of decreasing it (which gives worse results). This gives the following dynamics after discretization\begin{equation*}
\theta_{i} = \theta_{i-1} - \frac{\epsilon}{2}\nabla_{\theta} \widehat{\mathcal{U}}(\theta_i) + \zeta_i , \quad i = 1, \dots, R,
\label{eq:SGLDDynamics}
\end{equation*}
where $\widehat{\mathcal{U}}(\theta) = -\nabla_{\theta} \widehat{\ell}_{m}(\theta) - \log p_{\Theta}(\theta)$ with $\nabla_{\theta} \widehat{\ell}_{m}(\theta)$ in \eqref{eq:difference_estimator_grad_logl} and $\zeta_i \sim \mathcal{N}(0, \epsilon I)$. 

Following \cite{chen2014stochastic},  we implement SG-HMC using a discretized dynamics with momentum $\vec{p}$ with covariance matrix $M$ of the form
\begin{eqnarray*}
\theta_{l} & = & \theta_{l-1} + \epsilon M^{-1}\vec{p}_{l-1}  \\
\vec{p}_{l} & = & \vec{p}_{l-1}  - \epsilon \nabla_{\theta}\widehat{\mathcal{U}}(\theta_l) - \epsilon C M^{-1} \vec{p}_{l-1}   + \zeta_l , \quad l = 1, \dots, L,
\end{eqnarray*}
where $\widehat{\mathcal{U}}(\theta) = -\nabla_{\theta} \widehat{\ell}_{m}(\theta) - \log p_{\Theta}(\theta)$ with $\nabla_{\theta} \widehat{\ell}_{m}(\theta)$ in \eqref{eq:difference_estimator_grad_logl}, $\zeta_l \sim \mathcal{N}(0, 2(C-\widehat{B})\epsilon)$. We set $\widehat{B}= 0$ \citep{chen2014stochastic} and $C=I$ \citep{ma2015complete}.

The algorithms are compared with respect to CT as defined in \eqref{eq:CT} when the likelihood estimator is non-negative, where $\mathrm{IF}$ is computed using the CODA package in R \citep{plummer2006coda}. For signed HMC-ECS, the CT is given by \eqref{eq:CTSigned} and we estimate $\tau$ by the fraction of positive signs. Note that CT does not take into account that some of the algorithms can give a substantially biased estimate of the posterior, and we assess the bias separately.

The Relative Computational Time (RCT) between algorithm $\mathcal{A}_1$ and $\mathcal{A}_2$ is defined as
\begin{equation}
\label{eq:RCT_measure}
\mathrm{RCT}_{\mathcal{A}_1, \mathcal{A}_2} = \frac{\mathrm{CT}_{\mathcal{A}_2}}{\mathrm{CT}_{\mathcal{A}_1}}.
\end{equation}

\subsection{Tuning and settings of our algorithms}\label{subsec:tuning_our_algs}
We first outline how our algorithms are tuned. These settings are used for all the experiments unless stated otherwise.

The choice of the positive-definite mass matrix $M$ in \eqref{eq:energies} is crucial for the performance of any $\mathrm{HMC}$ type algorithm: an $M$ that closely resembles the covariance of the posterior facilitates sampling, especially when $\theta$ is highly correlated in the posterior \citep{neal2011mcmc, betancourt2017conceptual}. In logistic regression, we set $M = -\Sigma^{-1}(\theta^{\star})$, where $\Sigma(\theta^{\star})$ is the Hessian of the log posterior evaluated at some $\theta^{\star}$. We initialize $M = I$ and, during a burn-in period of $1{,}000$ iterations, update $M$ every $200$ iterations based on the new $\theta^\star$. We use the same tuning of $M$ in $\mathrm{HMC\mbox{-}ECS}$ when updating $\theta$ and $\vec{p}$ conditional on the data subsample $u$. However, since it is impractical to compute the Hessian of the conditional posterior at each iteration, we use the full dataset when evaluating $M$ (and include the cost in the CT), which performs well in practice although we stress that it is not optimal.

To select the step size $\epsilon$ in the leapfrog integrator, we utilize the dual averaging approach of \cite{hoffman2014no}, which requires a predetermined trajectory length $\epsilon L$. We find that $\epsilon L = 1.2$ is sensible for our examples. The dual averaging algorithm uses this trajectory length and adaptively changes $\epsilon$ during the burn-in period in order to achieve a desired level of acceptance rate $\delta$. We follow \cite{hoffman2014no} and set $\delta = 0.8$. The tuning for the logistic regression is relatively simple since $\Sigma$ can be computed analytically, which is useful for both setting $M$ and $\epsilon$. More complex models are harder to tune, but this is unlikely to influence the comparisons between $\mathrm{HMC\mbox{-}ECS}$ and the other algorithms here, which is our primary concern. We stress that these tuning issues are inherent to HMC itself and not due to our subsampling approach. This strategy gives $\epsilon = 0.2$ and $L = 6$ for our algorithm which is the default in our experiments unless otherwise stated.

In $\mathrm{HMC\mbox{-}ECS}$ we generate subsamples with a correlation of $\rho=0.99$. The subsample size $m$ in the perturbed $\mathrm{HMC\mbox{-}ECS}$ approach is set according to the guidelines in \cite{quiroz2018speeding}, see Section \ref{sec:Approx_HMCwG}. In the signed $\mathrm{HMC\mbox{-}ECS}$ approach, we follow \cite{quiroz2016exact} who set the mini-batch size $m = 30$ and set $\lambda$ optimally to minimize CT according to the formulas in \cite{quiroz2016exact}. For our examples, $\lambda = 100$ and  $\lambda = 200$, for the HIGGS and bankruptcy datasets, respectively.

\subsection{Tuning and settings of the competing algorithms}\label{subsec:tuning_other_algs}
We run full data HMC using the tuning strategy for $\epsilon$ and $M$ outlined in Section \ref{subsec:tuning_our_algs} and use $M = \Sigma^{-1}(\theta^\star)$ with $\theta^\star$ as the posterior mean in SG-HMC. This favours SG-HMC over HMC-ECS, which needs to learn $\theta^\star$ as the algorithm progresses. 

We are unaware of any tuning strategies for setting $\epsilon$ in SGLD and SG-HMC and therefore use trial and error. We find that the value of $\epsilon$ depends on which dataset or algorithm was considered; see Sections \ref{subsec:ResultsHiggs} and \ref{subsec:ResultsBankruptcy}. For SG-HMC we also need to set the number of steps $L$ and we explore two choices. First, notice that HMC-ECS uses the trajectory length $\epsilon L=1.2$. We thus set $L=1.2/\epsilon$ so that SG-HMC can traverse the space as swiftly as HMC-ECS. We also compare with SG-HMC using a value of $L$ which gives the same number of likelihood and gradient evaluations as HMC-ECS. For SGLD, we run the algorithm for $R$ iterations that correspond to the number of gradient evaluations used post burn-in in HMC-ECS. For example, $R = 12{,}000$ iterations if $L = 6$ and HMC-ECS performs $2,000$ iterations.

Finally, we use the same $m$ as HMC-ECS for SGLD and SG-HMC, as it is outside the scope of this paper to derive optimality results for those algorithms.

\subsection{Results for the HIGGS data}\label{subsec:ResultsHiggs}

 \cite{baldi2014searching} use the HIGGS dataset, which contains 11 million observations, with a binary response \emph{detected particle} predicted by 29 covariates. We use 10.5 millions observations for training and $500{,}000$ for testing. 

Unless stated otherwise, we start all algorithms at a $\theta^\star$ obtained as the posterior mean from running HMC on 1\% of a randomly chosen subset of the data. This $\theta^\star$ is also used to initialize the control variates. We first run the algorithms using a full mass matrix $M$, chosen as explained above. The subsample size was set to $m=1,300$ for all methods. For SG-HMC and SGLD, $\epsilon=0.06$ and $\epsilon=0.000001$ are used, respectively. The resulting $L$ is $20$ for SG-HMC.

Table \ref{tab:TableRCT_HIGGS} displays the CT for each algorithm compared to the perturbed $\mathrm{HMC\mbox{-}ECS}$ algorithm. The table shows the minimum, median and maximum RCT across all parameters. The best algorithm is perturbed $\mathrm{HMC\mbox{-}ECS}$ closely followed by signed $\mathrm{HMC\mbox{-}ECS}$, both have RCTs that are roughly three times better than the best competitor $\mathrm{SG\mbox{-}HMC}$. Although our metrics do not allow for direct comparison between biased (our approaches, SG-HMC and SGLD) methods and unbiased methods (HMC), we note that implementing HMC using the full data in this example is, for the perturbed and exact approaches respectively, 642.8 and 554.1 times more expensive in terms of posterior density and gradient evaluations.

	\begin{table}
		\centering
		\fontsize{11}{11}\selectfont
		\begin{tabular}{lrrr} \toprule
			& & &\\
			RCT  & Signed $\mathrm{HMC\mbox{-}ECS}$ & $\mathrm{SG\mbox{-}HMC}$  & $\mathrm{SGLD}$\\ [5pt]\midrule
			& & \\
			min  & 0.8 &2.43 & 3.58 \\ [5pt]
			median  & 1.15 &2.97 &12.46   \\[5pt]
			max  & 1.95  & 4.08 & 326.80\\[5pt]
			\bottomrule
		\end{tabular}	
		\vspace{3mm}
		\caption{HIGGS data. Relative computational time compared to perturbed HMC-ECS. The computational cost is (number of likelihood/gradient evaluations)$\times$IF. For HMC-ECS and signed HMC-ECS the cost is computed for the entire run including training and warmup period. The RCT for the stochastic gradient methods are based on post-burnin iterations only.}
		\label{tab:TableRCT_HIGGS}
	\end{table}
	
	\begin{figure}
	\centering
	\includegraphics[width=0.85\linewidth]{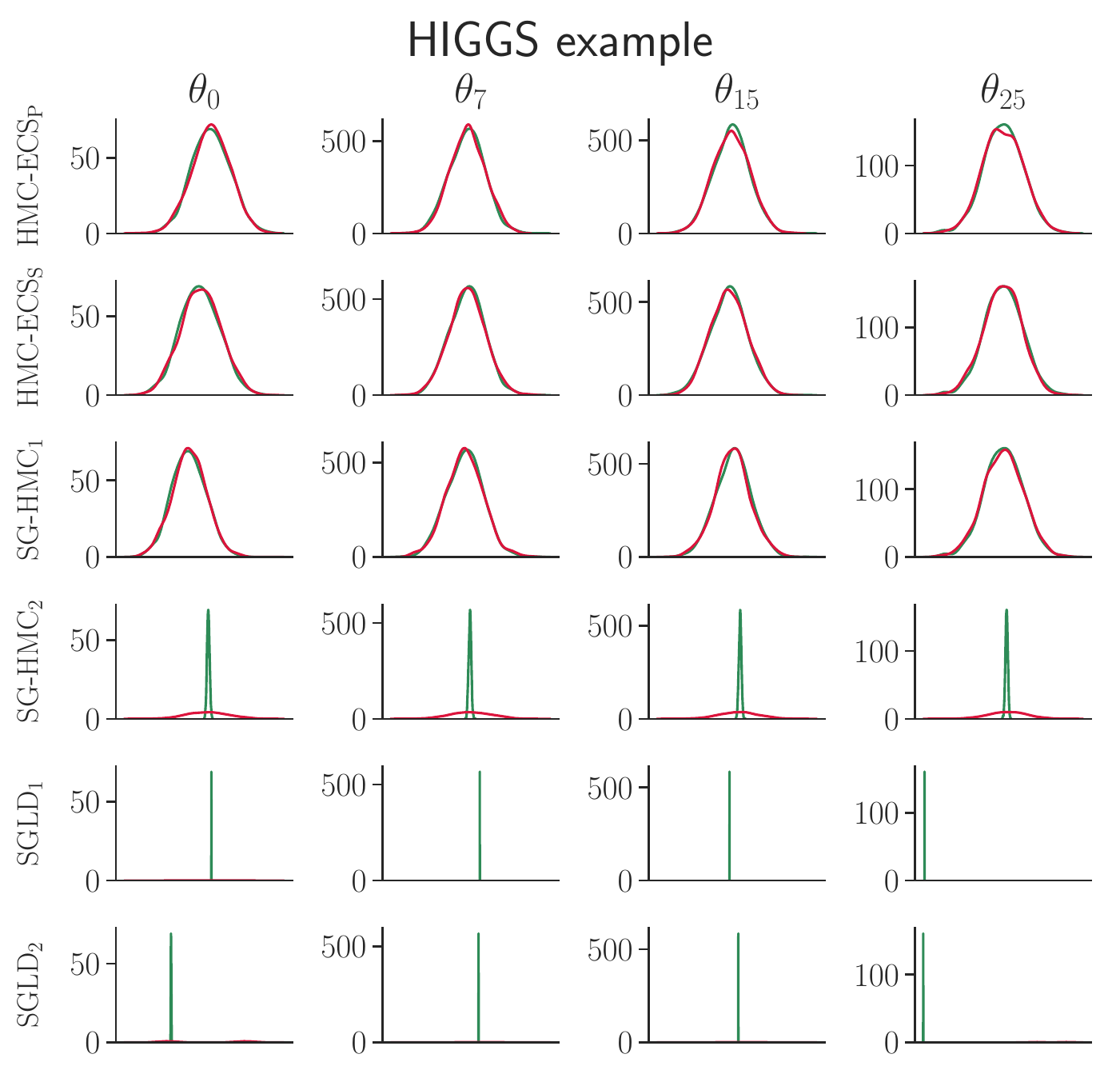}
	\caption{HIGGS data. The figure shows kernel density estimates of the posterior from the compared subsampling algorithms for four randomly selected parameters (green lines) and the corresponding posterior from HMC on the full dataset (red lines). All algorithms use the negative inverse Hessian from all the data as mass matrix. $\text{HMC\mbox{-}ECS}_\mathrm{P}$ and $\text{HMC\mbox{-}ECS}_\mathrm{S}$ denote, respectively, the perturbed and signed $\text{HMC\mbox{-}ECS}_\mathrm{S}$.  For SGLD and SG-HMC, subscript 1 refers to the second order control variate and subscript 2 refers to the version without control variates.}
	\label{fig:KDE_HIGGS_Hessian}
\end{figure}

\begin{figure}
	\centering
	\includegraphics[width=0.65\linewidth]{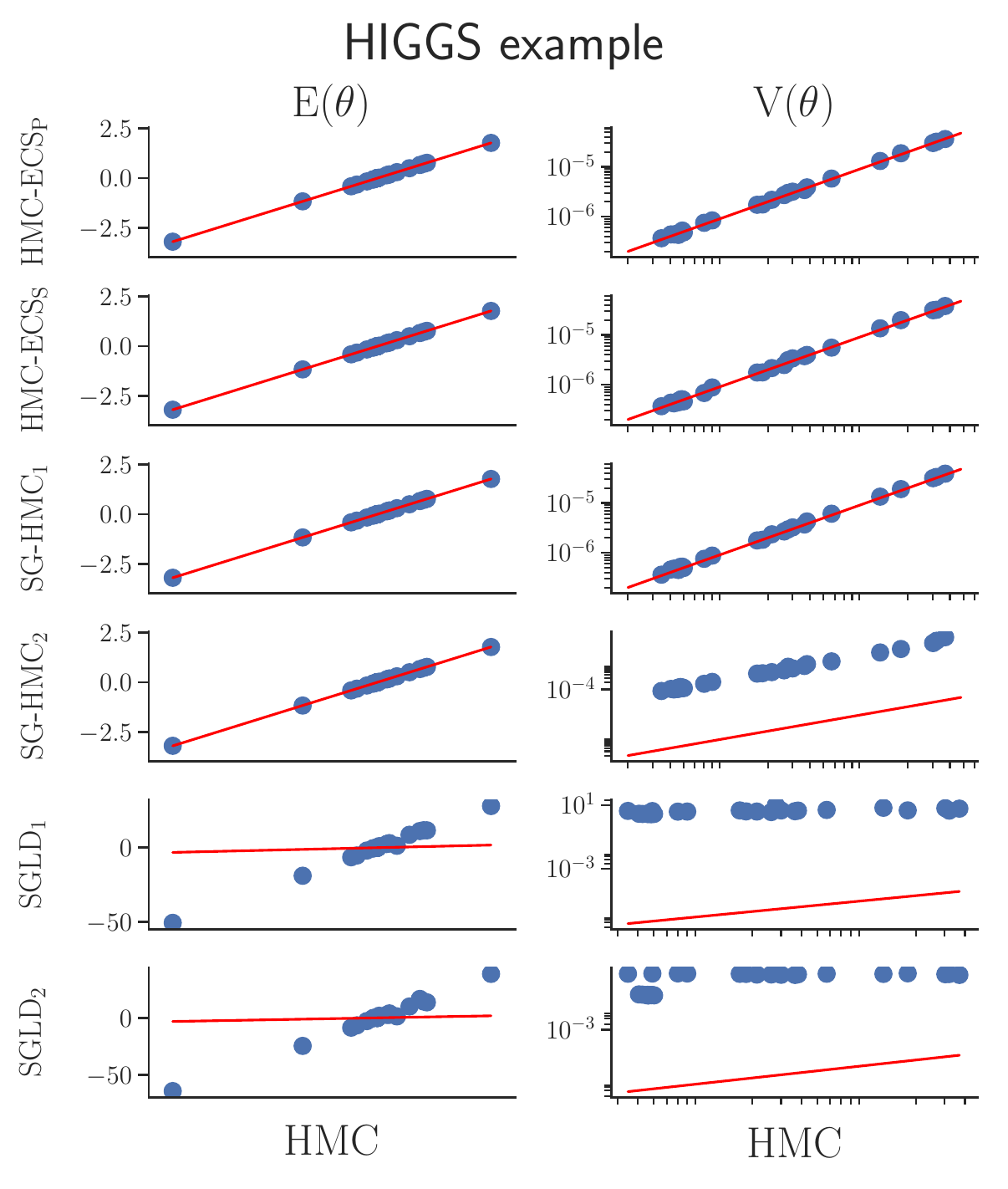}
	\caption{HIGGS data. The figure plots the estimated posterior mean and variance for all parameters from the subsampling algorithm against their true values obtained by HMC on the full dataset. All algorithms use the negative inverse Hessian from all the data as mass matrix. $\text{HMC\mbox{-}ECS}_\mathrm{P}$ and $\text{HMC\mbox{-}ECS}_\mathrm{S}$ denote, respectively, the perturbed and signed $\text{HMC\mbox{-}ECS}_\mathrm{S}$. For SGLD and SG-HMC, subscript 1 refers to the second order control variate and subscript 2 refers to the version without control variates.}
	\label{fig:MeanVar_HIGGS_Hessian}
\end{figure}
	
RCT does not take the bias into account, however. Figure \ref{fig:KDE_HIGGS_Hessian} displays kernel density estimates of the marginal posterior of four randomly selected parameters, and Figure \ref{fig:MeanVar_HIGGS_Hessian} plots the posterior mean and variance for all the parameters from each algorithm against the true posterior mean and variance obtained from HMC based on the full dataset. Figures \ref{fig:KDE_HIGGS_Hessian} and \ref{fig:MeanVar_HIGGS_Hessian} clearly show that both HMC-ECS algorithms and SG-HMC do a very good job in approximating the posterior, while SGLD gives biased estimates. We have also added the results for SG-HMC and SGLD without control variates as variance reduction is not always optimal for these algorithms \citep{pmlr-v80-chatterji18a}. In this example, control variates are indeed helpful, except for SGLD which does not provide an accurate approximation regardless of which control variate is used.

\begin{table}
	\centering
	\fontsize{11}{11}\selectfont
	\begin{tabular}{lrr} \toprule
		& Perturbed HMC-ECS & Signed HMC-ECS\\ [5pt]\midrule
		$\alpha_{\theta, \vec{p}}$  & 0.980 & 0.979 \\ [5pt]
		$\alpha_u$  & 1&0.993     \\[5pt]
		$L$  & 6 & 6 \\[5pt]
		$\mathrm{IF}$  &2.185 & 2.192   \\[5pt]
		$\mathrm{ESS}$  &927 & 922  \\ [5pt] 
		$\widehat{\tau}$  &  & 1 \\ [5pt] 
		$100(m/n)$  &0.012 &0.029 \\ [5pt]\bottomrule
	\end{tabular}	
	\vspace{3mm}
	\caption{HIGGS data. Summary of settings and efficiencies of HMC-ECS. The table shows the average acceptance probabilities (as a benchmark HMC has 0.980) in the post burn-in period for the two Gibbs steps, the number of steps $L$ in the integrator used to obtain a predetermined trajectory length $\epsilon L = 1.2$, the average Inefficiency Factor (IF) (as a benchmark HMC has 2.084), the Effective Sample Size $\mathrm{ESS} =R/ \mathrm{IF}$, the estimated probability of a positive likelihood estimator $\tau$ and the percentage of data used by each of the algorithms.}
	\label{tab:SummaryHiggs}
\end{table}

\begin{figure}
	\centering
	\includegraphics[width=0.8\linewidth]{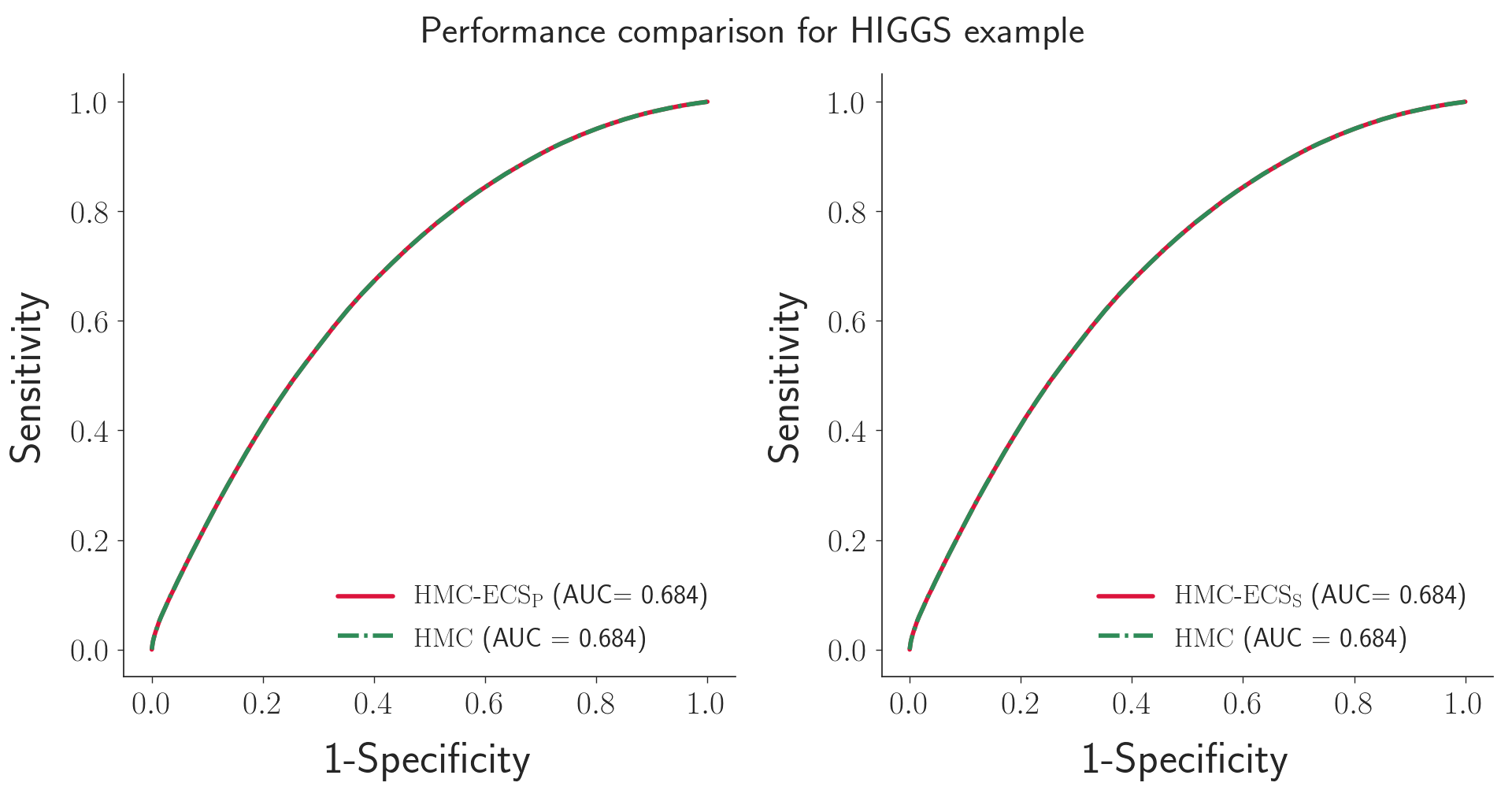}
	\caption{Prediction performance for the HIGGS data. The figure shows the Receiver Operating Characteristic (ROC) curves for the $500,000$ test observations with $\mathrm{HMC}$ and $\mathrm{HMC\mbox{-}ECS_\mathrm{P}}$ (left panel, perturbed $\mathrm{HMC\mbox{-}ECS}$) and $\mathrm{HMC\mbox{-}ECS_\mathrm{S}}$ (right panel, signed $\mathrm{HMC\mbox{-}ECS}$).}
	\label{fig:roc}
\end{figure}

We conclude this example by demonstrating that HMC-ECS can safely be used for obtaining the predictive distribution. Figure \ref{fig:roc} shows that the Receiver Operating Characteristic (ROC) curve for the $500,000$ test observations obtained with either of the two HMC-ECS algorithms are indistinguishable from the ROC curve obtained with HMC on the full dataset.

\subsection{Results for the Bankruptcy data}\label{subsec:ResultsBankruptcy}

This dataset contains annual observations on the bankruptcy status (binary  $y$) of Swedish firms in the time period 1991-2008. We follow \citet{giordani2014taking} and model the log odds of the firm failure probability as a non-linear function of six firm-specific financial ratios and two macroeconomic variables using an additive spline model. \citet{giordani2014taking} estimate the model using frequentist methods. We use a Bayesian approach with an $81$ dimensional posterior distribution (an intercept and $10$ basis spline functions for each covariate) given the $n=4{,}748{,}089$ firm-year observations.

Experimentation shows that $\epsilon =0.02$ is a sensible choice for SG-HMC. Using the same trajectory length as in HMC-ECS gives $L = 60$ for SG-HMC. We also compare with SG-HMC using $L = 7$ for which SG-HMC has the same number of gradient evaluations as HMC-ECS has gradient and likelihood evaluations. For SGLD, $\epsilon =0.00002$ is a sensible choice.

The subsample size for perturbed HMC-ECS was initially set to the optimal $m=62{,}000$ following the guidelines in \cite{quiroz2018speeding} based on the initial value for $\theta^\star$. We then ran perturbed HMC-ECS for $100$ iterations to obtain a better $\theta^\star$ and recalibrated to the now optimal $m=1{,}000$ for this improved $\theta^\star$. This improved $\theta^\star$ is also used to tune $\lambda$ in signed HMC-ECS, as explained in Section \ref{sec:Exact_HMCwG}. All iterations used for tuning are included in the computational cost.

	\begin{table}
		\centering
		\fontsize{11}{11}\selectfont
		\begin{tabular}{lrrrr} \toprule
			& & &\\
			RCT  & $\mathrm{HMC\mbox{-}ECS_{S}}$ & $\mathrm{SG\mbox{-}HMC_1}$ & $\mathrm{SG\mbox{-}HMC_2}$  & $\mathrm{SGLD}$ \\ [5pt]\midrule
			& & \\
			min  &  1.2 &6.8 &48.6 & 53.9\\ [5pt]
			median   & 1.6 &9.5 &100.2 & 230.1  \\[5pt]
			max    & 2.6  & 682.3 & 246.7& 2784.2\\[5pt]
			\bottomrule
		\end{tabular}	
		\vspace{3mm}
		\caption{Bankruptcy data. Relative computational time compared to perturbed HMC-ECS. For the two HMC-ECS algorithms, the cost is computed for the entire run including training and warmup period. The RCT with respect to stochastic gradient methods are based on post-burnin iteration only. $\mathrm{HMC\mbox{-}ECS}_{S}$ is the signed $\mathrm{HMC\mbox{-}ECS}$. $\mathrm{SG\mbox{-}HMC_1}$ and $\mathrm{SG\mbox{-}HMC_2}$ denote, respectively, the $\mathrm{SG\mbox{-}HMC}$ with $L=60$ and $L = 7$ leapfrog steps.} 
		\label{tab:TableRCT_Bankruptcy}
	\end{table}

\begin{figure}
	\centering
	\includegraphics[width=0.85\linewidth]{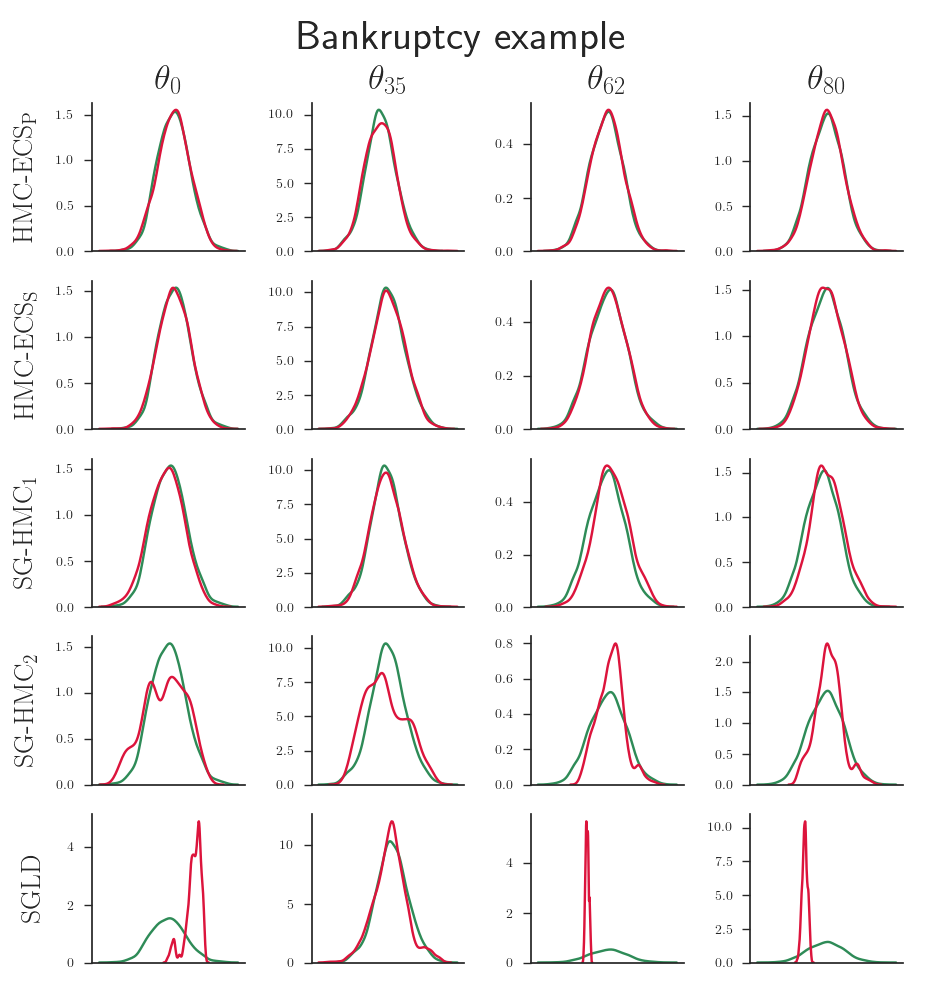}
	\caption{Bankruptcy data. The figure shows kernel density estimates of the posterior from the compared subsampling algorithms for four randomly selected parameters (green lines) and the corresponding posterior from HMC on the full dataset (red lines). $\text{HMC\mbox{-}ECS}_\mathrm{P}$ and $\text{HMC\mbox{-}ECS}_\mathrm{S}$ denote, respectively, the perturbed and signed $\text{HMC\mbox{-}ECS}_\mathrm{S}$. $\mathrm{SG\mbox{-}HMC_1}$ and $\mathrm{SG\mbox{-}HMC_2}$ denote, respectively, the $\mathrm{SG\mbox{-}HMC}$ with $L=60$ and $L = 7$ leapfrog steps.}
	\label{fig:KDE_Bankruptcy_Hessian}
\end{figure}

\begin{figure}
	\centering
	\includegraphics[width=0.8\linewidth]{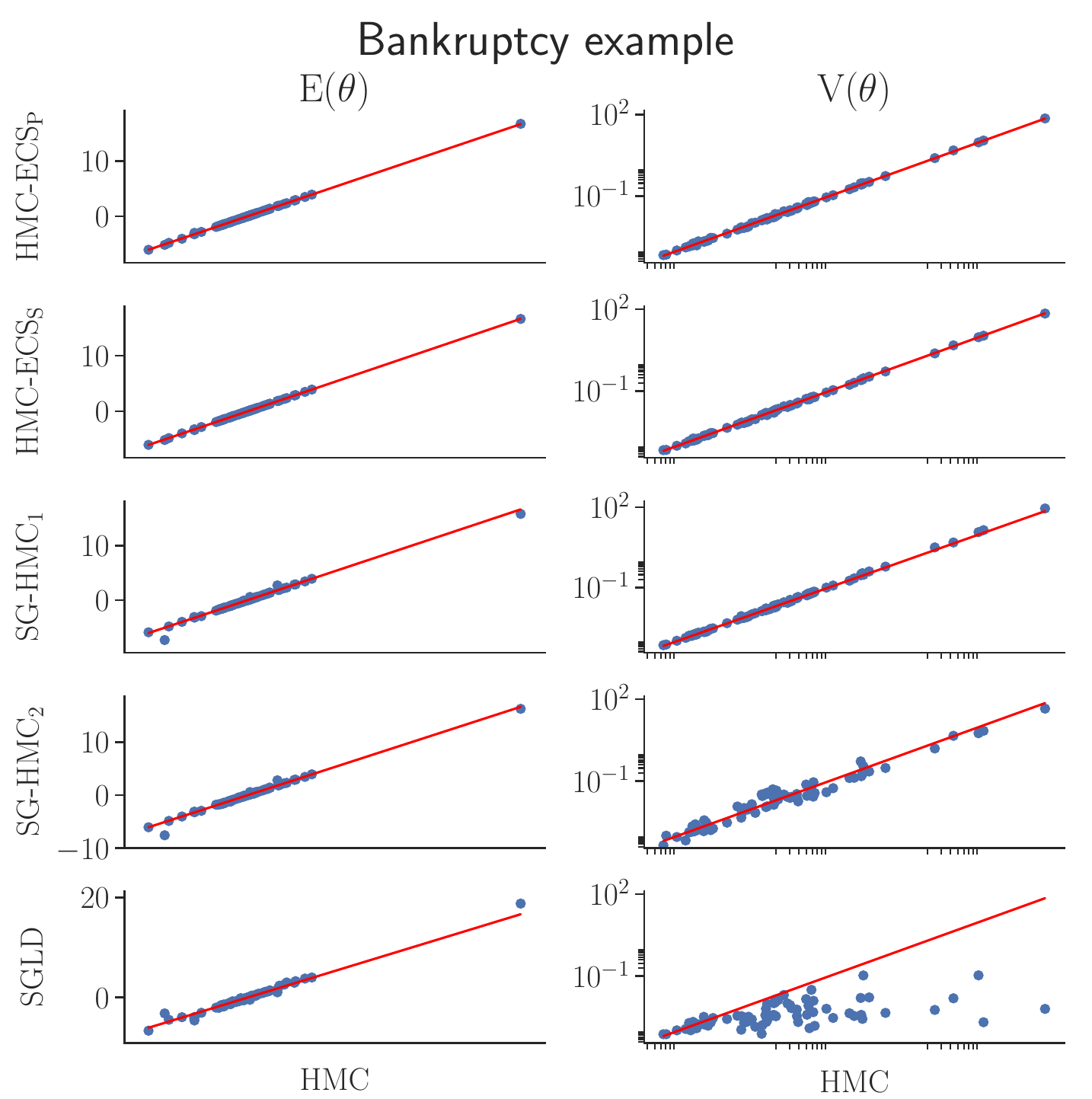}
	\caption{Bankruptcy data. The figure plots the estimated posterior mean and variance for all parameters from the subsampling algorithm against their true values obtained by HMC on the full dataset. $\text{HMC\mbox{-}ECS}_\mathrm{P}$ and $\text{HMC\mbox{-}ECS}_\mathrm{S}$ denote, respectively, the perturbed and signed $\text{HMC\mbox{-}ECS}_\mathrm{S}$. $\mathrm{SG\mbox{-}HMC_1}$ and $\mathrm{SG\mbox{-}HMC_2}$ denote, respectively, the $\mathrm{SG\mbox{-}HMC}$ with $L=60$ and $L = 7$ leapfrog steps.}
	\label{fig:MeanVar_Bankruptcy_Hessian}
\end{figure}

\begin{table}
	\centering
	\fontsize{11}{11}\selectfont
	\begin{tabular}{lrr} \toprule
		 & Perturbed HMC-ECS & Signed HMC-ECS\\ [5pt]\midrule
        $\alpha_{\theta, \vec{p}}$   &0.967 & 0.962\\ [5pt]
		$\alpha_u$  &0.994 &0.964   \\[5pt]
		$L$  & 6 & 6\\[5pt]
		$\mathrm{IF}$  &2.202 & 2.31 \\[5pt]
		$\mathrm{ESS}$ &912 & 871 \\ [5pt] 
		$\widehat{\tau}$  & & 1\\ [5pt] 
		$100(m/n)$  &0.021 &0.126 \\ [5pt]\bottomrule
	\end{tabular}	
	\vspace{3mm}
	\caption{Bankruptcy data. Summary of settings and efficiencies of HMC and HMC-ECS. The table shows the average acceptance probabilities (as a benchmark HMC has 0.966) in the post burn-in period for the two Gibbs steps, the number of steps $L$ in the integrator used to obtain a predetermined trajectory length $\epsilon L = 1.2$, the average Ineffiency Factor (IF) (as a benchmark HMC has 2.195), the Effective Sample Size $\mathrm{ESS} =R/ \mathrm{IF}$, the estimated probability of a positive likelihood estimator $\tau$ and the percentage of data used by each of the algorithms.}
	\label{tab:SummaryBankruptcy}
\end{table}

Figures \ref{fig:KDE_Bankruptcy_Hessian} and \ref{fig:MeanVar_Bankruptcy_Hessian} clearly show that the two HMC-ECS algorithms give better posterior approximations. The figures also show that SG-HMC with $L=60$ leapfrog steps performs reasonably well in terms of accuracy and much better than SGLD, although some expectation estimates are biased. SG-HMC with $L=7$ has substantially higher inefficiency factors and these figures show the degraded accuracy of the algorithm.

\begin{figure}
	\centering
	\includegraphics[width=0.9\linewidth]{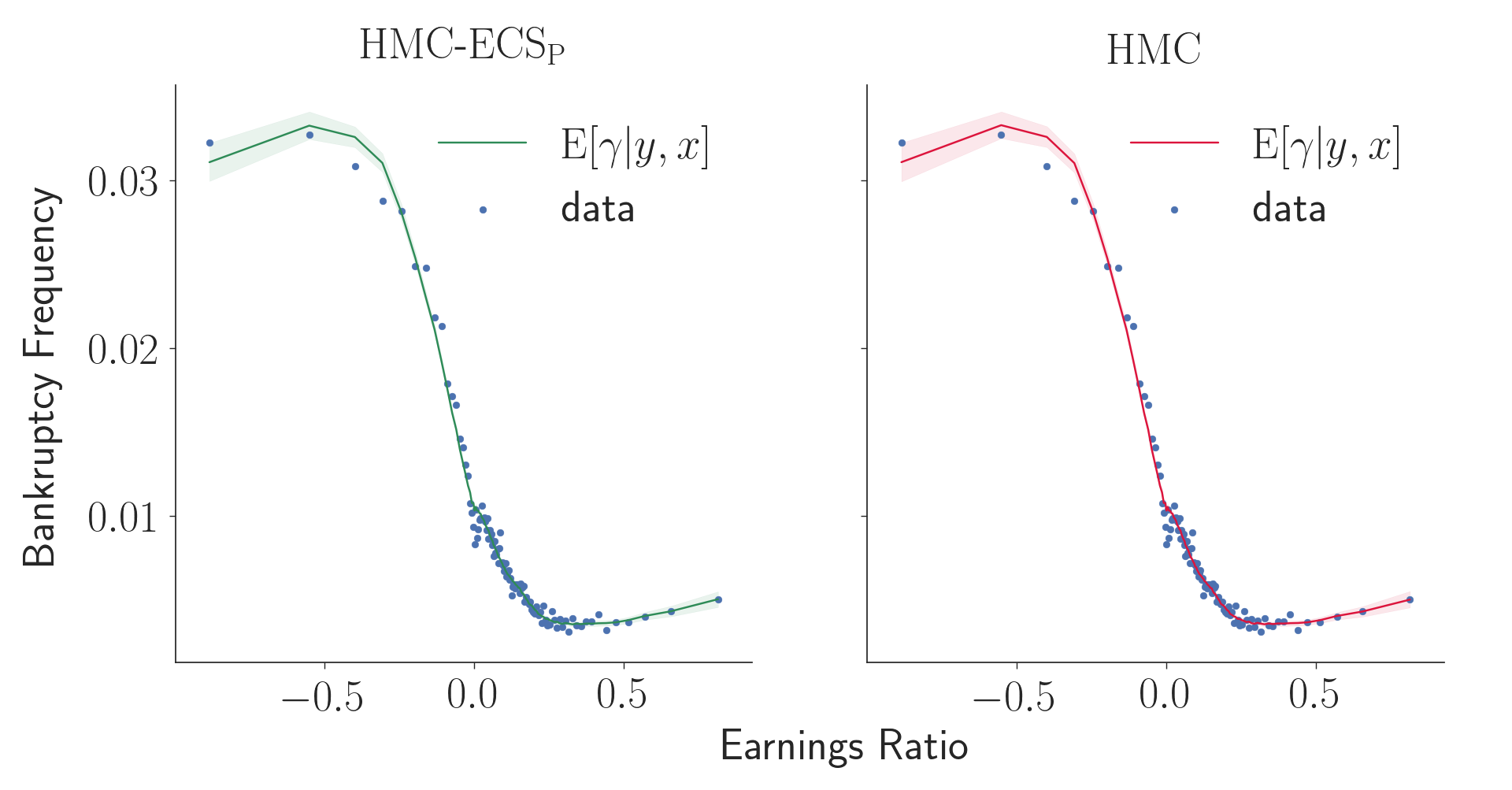}
	\caption{Realized and predicted bankruptcy probabilities as a function of the covariate \emph{Earnings ratio} from the perturbed $\mathrm{HMC\mbox{-}ECS}$ (left panel) and  $\mathrm{HMC}$ (right panel). 
Realized mean bankruptcy probabilities (blue dots) are computed by dividing the data into $100$ equally sized groups based on the earnings ratio variable and estimating the bankruptcy probability by the fraction of bankrupt firms in each group. The solid line is the predictive mean and the shaded regions are point wise $90$\% equal tail posterior credible intervals.} 
	\label{fig:clustermeanpx1}
\end{figure}

Figure \ref{fig:clustermeanpx1} shows the probability of bankruptcy for the fitted model and the empirical bankruptcy frequencies as a function of one of the covariates, Earnings ratio, with details in the caption. The posterior mean and posterior predictive intervals obtained by $\mathrm{HMC}$ and the perturbed $\mathrm{HMC\mbox{-}ECS}$ are indistinguishable.

Finally, we note that implementing HMC using the full data in this example is, for the perturbed and exact approach respectively, 478.7 and 311.5 times more expensive in terms of posterior density and gradient evaluations.

\subsection{Scalability of HMC-ECS}
\label{subsec:scalability}
\cite{beskos2013optimal} show that, in an optimally tuned HMC algorithm, the step size $\epsilon$ needs to be scaled as $O(d^{-1/4})$ to keep the acceptance probability constant as the dimension $d$ increases. This is more favorable than the rate $O(d^{-1/3})$ of Langevin Monte Carlo \citep{roberts1998optimal}. We have argued that since our algorithm is performing a HMC step using a Hamiltonian based on a subset of the data it should scale with dimension similarly to HMC. 

\begin{figure}
	\centering
	\includegraphics[width=0.9\linewidth]{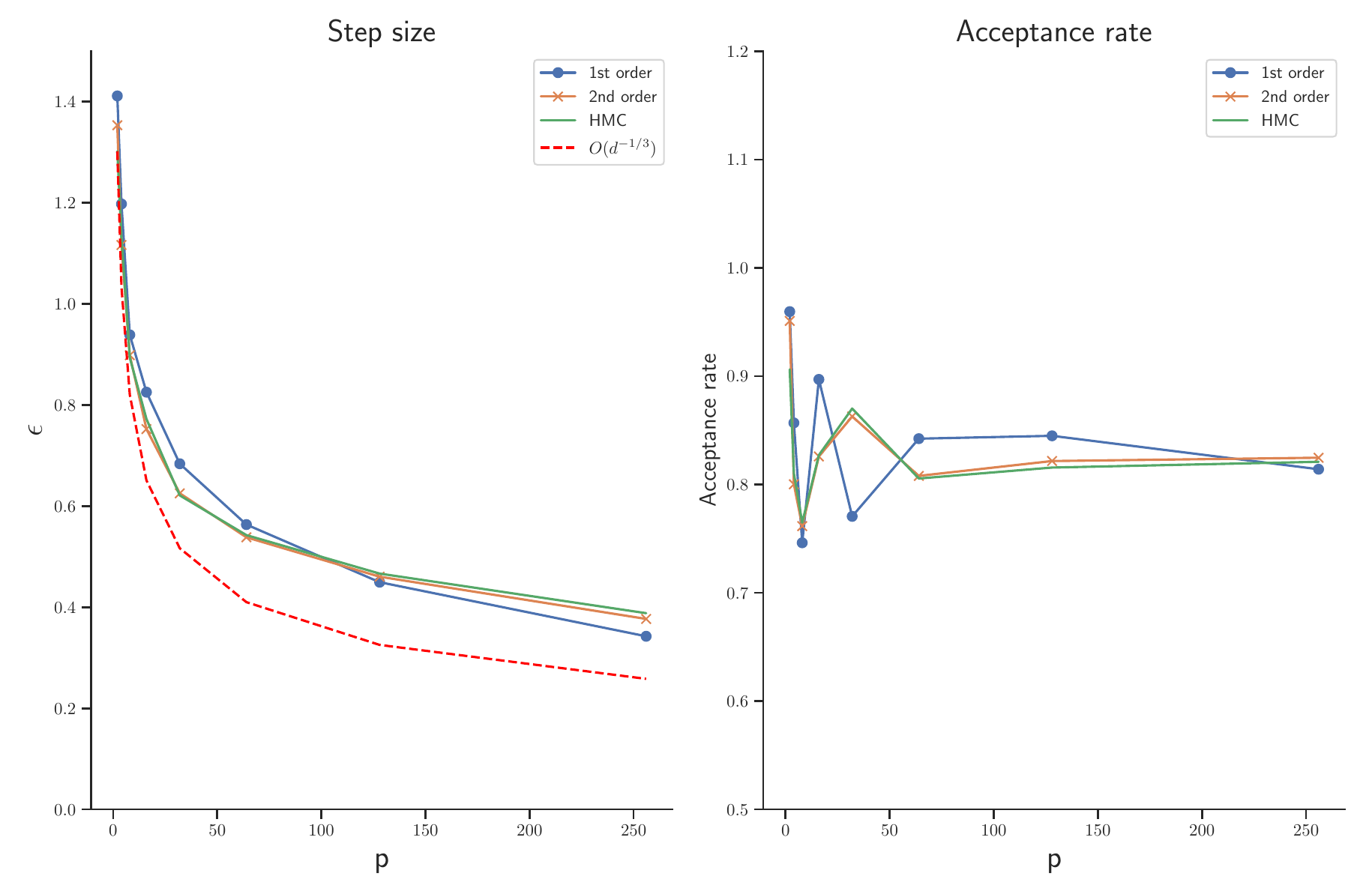}
	\caption{Empirical illustration of the scaling of HMC-ECS, see Section \ref{subsec:scalability} for the experimental settings. The left figure shows the optimal step size as a function of the dimension. HMC ($O(d^{-1/4})$) and Langevin ($O(d^{-1/3})$) are plotted for reference. The right figure shows the acceptance rate targeted for optimality. The IF of our methods are close to 1, regardless of if the 1st or 2nd control variate is used. The variance of the log-likelihood estimator for the second order control variate is nearly zero (quadratic target) and for the first order control variate is kept around 1 \citep{pitt2012some} by selecting $m$ appropriately.} 
	\label{fig:scaling_HMC_ECS}
\end{figure}

We set out to test this hypothesis empirically as follows. First, we consider a sequence of $d$, obtained as $d=2^h$, $h=1,\dots, 8$ and obtain eight simulated datasets with $n=10,000$ each. For each $d$, we run the dual averaging algorithm as described in Section \ref{subsec:tuning_our_algs} to find the optimal $\epsilon$and check if $\epsilon =O(d^{-1/4})$ is reasonable. In agreement with \cite{beskos2013optimal} we set $M$ optimally from the curvature of the conditional target posterior. This is easily achieved by considering a Gaussian regression model where we set the prior $p_{\Theta}(\theta)=\mathcal{N}(0, 5^2I_d)$, such that the optimal $M$ for the conditional target is, assuming the bias-correction term to be negligible, $$M=\frac{n}{m}\sum_{i=1}^m \mathbf{X}^\top_{u_i}\mathbf{X}_{u_i} + \frac{1}{5^2},$$
where $\mathbf{X}_{u_i}$ denotes the $u_i$th row of the design matrix. We note that, since the Gaussian model is quadratic in its log-density, the second order control variate will yield a perfect fit, i.e. the variance of $\widehat{l}_m(\theta)$ is zero. Thus, we also experiment with a first order control variate. We scale $m$ to maintain the variance around 1 \citep{pitt2012some}. Figure \ref{fig:scaling_HMC_ECS} shows the results and we deduce that the algorithm does indeed maintain scalability.

\subsection{Limitations of subsampling HMC}
\label{subsec:Limitations}
\begin{figure}
	\centering
	\includegraphics[width=0.9\linewidth]{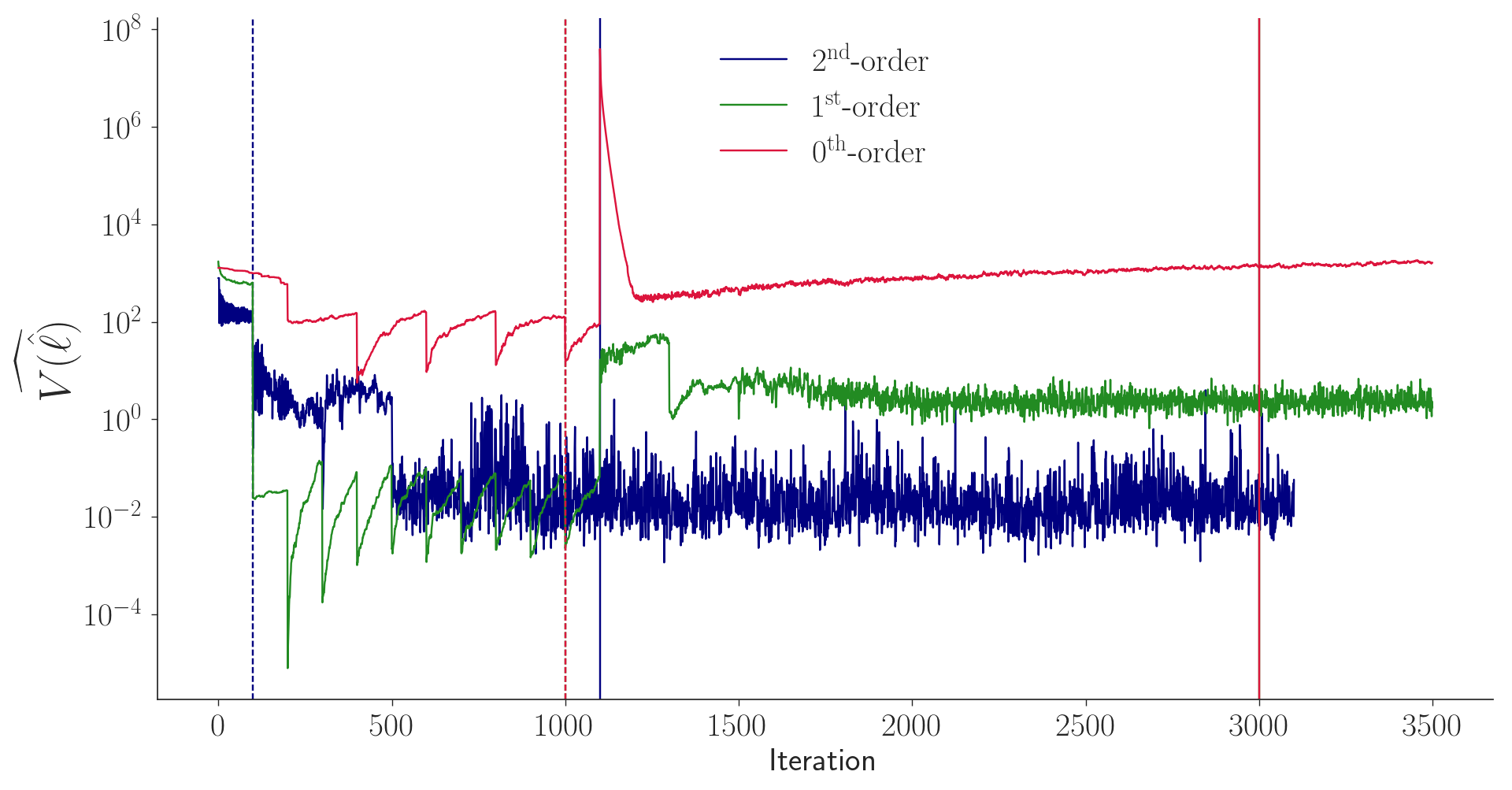}
	\caption{Variance of $\widehat{l}_m$. The figure shows the estimated variance as a function of the iterations for three different orders in the Taylor expansion for the control variates. The dashed vertical lines correspond to the end of the training period used the solid vertical line corresponds to the end of the burn-in period.}
	\label{fig:variances_allTaylor}
\end{figure}

\begin{figure}
	\centering
	\includegraphics[width=0.8\linewidth]{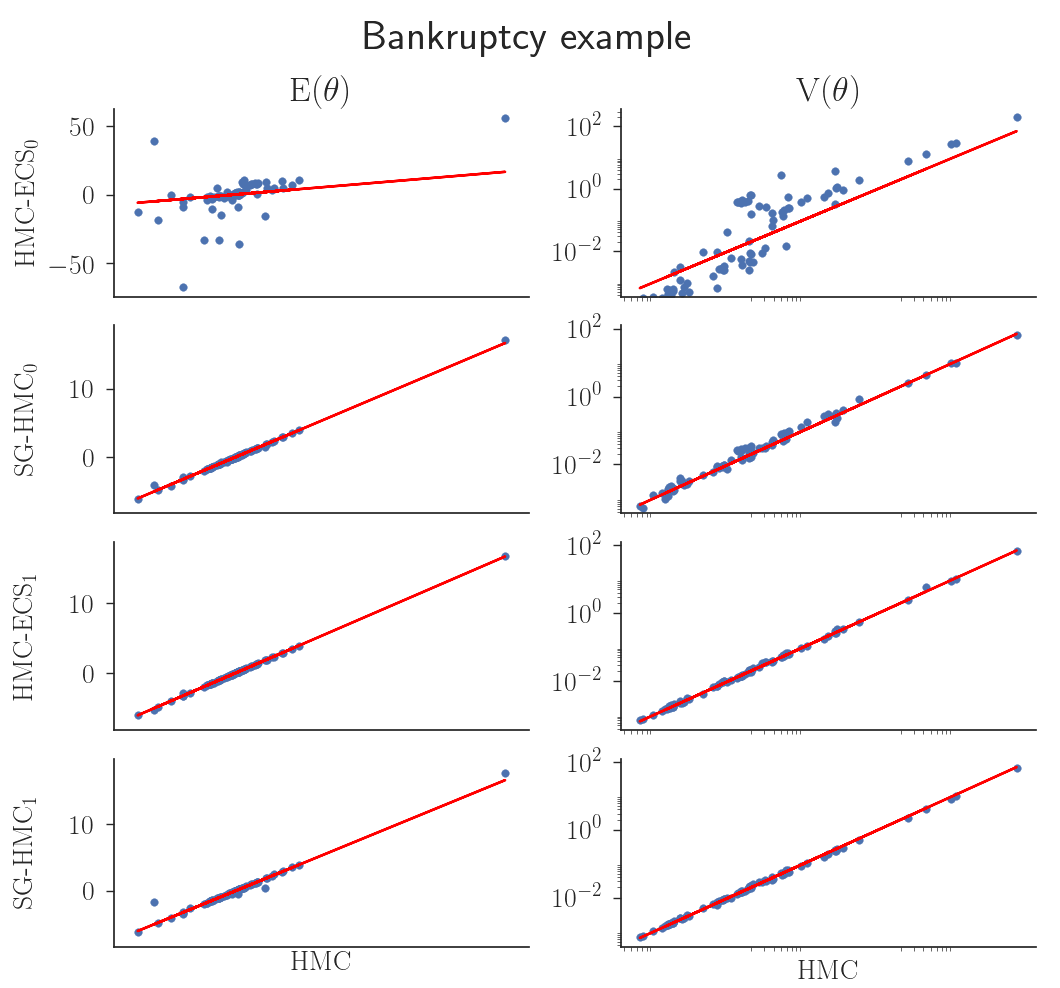}
	\caption{Results for the zeroth and first order control variate. 
	The figure shows the estimates of posterior expectations and posterior variances using the zeroth order control variate (upper panel) and first order control variate (lower panel). All comparisons are versus HMC which represents the ground truth.}
	\label{fig:results_0th_and_1th_order_control_variate}
\end{figure}

Variance reduction by control variates is crucial in any subsampling MCMC algorithm. This subsection explores the role of control variates by successively degrading the quality of the control variates by lowering the order of their Taylor approximation. Figure \ref{fig:variances_allTaylor} shows the estimated variance of $\widehat{l}_m(\theta)$ as a function of the iterates when the control variates are based on a Taylor series expansions of different orders. We note that the algorithm survives a substantial variance during the training iterations when applying the first order control variates, and this variance eventually settles down once a sensible  $\theta^\star$ is found. Figure \ref{fig:results_0th_and_1th_order_control_variate} shows that the expectation and variance estimates are very accurate for the first order control variates. Figures \ref{fig:variances_allTaylor} and \ref{fig:results_0th_and_1th_order_control_variate} also shows that control variates of zero order are too crude for HMC-ECS in this example. While it seems that the competing methods are more robust to the quality of the control variates, we again stress that all competitors are placed in the unrealistically favorable scenario of having $M$ equal to the inverse Hessian evaluated at the posterior mean.

The control variates \citep{bardenet2015markov} used in this paper are quadratic in $\theta$ centered around $\theta^\star$. This means that they are expected to work well for any model which has a log-density that is reasonably quadratic in $\theta$ in a neighborhood of $\theta^\star$. Examples are Poisson regression \citep{quiroz2018review} and Student-t regression \citep{quiroz2018speeding}. While the locally approximately quadratic feature is found in many models, there are clearly models in which this local approximation can be poor, for example deep neural nets. To find efficient control variates, which are also computationally feasible, in a class of complex models remains an open challenging problem. We stress however that this is a problem for all existing subsampling MCMC approaches, and any progress on improved control variates can be straightforwardly incorporated into HMC-ECS.

\section{Conclusions and Future Research\label{sec:Conclusions}}
We propose a method to speed up Bayesian inference while maintaining high sampling efficiency in moderately high-dimensional parameter spaces by combining data subsampling and Hamiltonian Monte Carlo such that the energy is conserved. We show how to implement the method using two estimators of the likelihood. The first implementation, which we refer to as perturbed HMC-ECS, produces iterates from a perturbed density that will get arbitrarily close to the true density, as measured by the total variation metric, at the rate $O(n^{-1}m^{-2})$. The second implementation, which we refer to as signed HMC-ECS, gives iterates which are then used in an importance sampling estimator to obtain a simulation consistent estimator of the expectation of any posterior functional. 

We apply the methods to simulate from the posterior distribution in two datasets, with $d=29$ and $d=81$ dimensions, respectively. Our two HMC-ECS algorithms perform highly accurate inference, comparable to  HMC without subsampling, but are computationally much faster.  This is a major step forward since \cite{bardenet2015markov} and \cite{quiroz2018speeding} demonstrate that most subsampling approaches cannot even beat standard MH without subsampling on toy examples with $d=2$ and highly redundant data. We also show that HMC-ECS is very competitive against SGLD and SG-HMC, both in terms of sampling efficiency and accuracy.

Control variates to reduce the variance of subsampling estimators are well known to be crucial for any subsampling MCMC algorithm. We use very efficient control variates based on a second order Taylor expansion in our applications, but explore the effects of less accurate control variates. We find that HMC-ECS still performs well with a cruder first order approximation, but that a Taylor approximation of order zero is too crude and gives a too large variance for HMC-ECS. 

Similarly to HMC, $\mathrm{HMC\mbox{-}ECS}$ is difficult to tune. Self-tuning algorithms such as the no-U-Turn sampler \citep{hoffman2014no} have been proposed for HMC and it would be interesting to see if our ideas can be applied there. It would also be interesting to consider Riemann Manifold HMC \citep{girolami2011riemann}, which has been demonstrated to be very effective when a high-dimensional posterior exhibits strong correlations. Scaling up such an algorithm opens up the possibility of simulating the posterior density of highly complex models with huge datasets. Finally, until recently, one of the limitations of HMC was its inability to cope with discrete parameters. \cite{nishimura2017discontinuous} overcomes this limitation and extending $\mathrm{HMC\mbox{-}ECS}$ in this direction would be an interesting undertaking.

\section{Acknowledgments}
We thank the action editor and two reviewers for constructive comments that improved both the content and presentation of the paper.

Matias Quiroz and Robert Kohn were partially supported
by Australian Research Council Center of Excellence grant CE140100049. Mattias Villani was partially supported by Swedish Foundation for Strategic
Research (Smart Systems: RIT 15-0097).
\bibliographystyle{apalike}
\addcontentsline{toc}{section}{\refname}\bibliography{references}


\end{document}